\documentclass[
  journal=pasa,
  manuscript=research-paper, 
  year=2020,
  volume=37,
]{cup-journal}

\usepackage{microtype,siunitx,booktabs}

\usepackage{graphicx}
\usepackage{amsmath}	
\usepackage{amssymb}	
\usepackage{multicol}       
\usepackage{bm}		
\usepackage{pdflscape}	
\usepackage{xurl}
\usepackage{hyperref} 
\usepackage[T1]{fontenc}
\usepackage{ae,aecompl}
\usepackage{xcolor} 
\usepackage{booktabs}
\usepackage{pifont}
\usepackage{nicematrix}
\newcommand{\cmark}{\ding{51}}%
\newcommand{\xmark}{\ding{55}}%

\title[Radio Galaxy Detection with Machine Learning]{RadioGalaxyNET: Dataset and Novel Computer Vision Algorithms for the Detection of Extended Radio Galaxies and Infrared Hosts}

\author{Nikhel Gupta$^{1}$}
\author{Zeeshan Hayder$^{2}$} 
\author{Ray P. Norris$^{3,4}$} 
\author{Minh Huynh$^{1,5}$}
\author{Lars Petersson$^{2}$}
\email[Nikhel Gupta]{Nikhel.Gupta@csiro.au}
\affiliation{
$^1$ CSIRO Space \& Astronomy, PO Box 1130, Bentley WA 6102, Australia \\
$^2$ CSIRO Data61, Black Mountain ACT 2601, Australia \\
$^3$ Western Sydney University, Locked Bag 1797, Penrith, NSW 2751, Australia \\
$^4$ CSIRO Space \& Astronomy, P.O. Box 76, Epping, NSW 1710, Australia \\
$^5$ International Centre for Radio Astronomy Research (ICRAR), M468, The University of Western Australia, 35 Stirling Highway, Crawley, WA 6009, Australia \\ 
}

\received {dd Mmm YYYY}
\revised  {dd Mmm YYYY}
\accepted {dd Mmm YYYY}
\published{dd Mmm YYYY}

\keywords{galaxies: active; galaxies: peculiar; radio continuum: galaxies; Galaxy: evolution; methods: data analysis} 

\usepackage{amsmath,amsfonts,amssymb}
\usepackage{color}
\usepackage{mathrsfs}
\usepackage{tabularx}
\usepackage{floatrow}
\usepackage{float}

\definecolor{ored}{rgb}{1.00,0.27,0.00}
\definecolor{mygreen}{rgb}{0.2,0.7,0.2}

\definecolor{Gray}{gray}{0.5}
\definecolor{LightCyan}{rgb}{0.88,1,1}

\def \BE{\begin{equation}}
\def \EE{\end{equation}}	
\def \BC{\begin{center}}
\def \EC{\end{center}}
\def \BEA{\begin{eqnarray}}
\def \EEA{\end{eqnarray}}

\def \SIGMA8{\sigma_{8}}

\usepackage{cellspace}
\begin{document}\sloppy\sloppypar\raggedbottom\frenchspacing


\begin{abstract}
Creating radio galaxy catalogues from next-generation deep surveys requires automated identification of associated components of extended sources and their corresponding infrared hosts. 
In this paper, we introduce RadioGalaxyNET, a multimodal dataset, and a suite of novel computer vision algorithms designed to automate the detection and localization of multi-component extended radio galaxies and their corresponding infrared hosts. 
The dataset comprises 4,155 instances of galaxies in 2,800 images with both radio and infrared channels. 
Each instance provides information about the extended radio galaxy class, its corresponding bounding box encompassing all components, the pixel-level segmentation mask, and the keypoint position of its corresponding infrared host galaxy. 
RadioGalaxyNET is the first dataset to include images from the highly sensitive Australian Square Kilometre Array Pathfinder (ASKAP) radio telescope, corresponding infrared images, and instance-level annotations for galaxy detection. 
We benchmark several object detection algorithms on the dataset and propose a novel multimodal approach to simultaneously detect radio galaxies and the positions of infrared hosts.
\end{abstract}


\section{Introduction}
\label{SEC:Intro}
Recent advancements in radio astronomy have enabled us to scan large areas of the sky in a short timescale while generating incredibly sensitive continuum images of the Universe.
This has created new possibilities for detecting millions of galaxies at radio wavelengths. 
For example, the ongoing Evolutionary Map of the Universe \citep[EMU;][]{norris21} survey, conducted using the Australian Square Kilometre Array Pathfinder \citep[ASKAP;][]{johnston07ASKAP,DeBoer09,hotan21} telescope, is projected to discover more than 40 million compact and extended galaxies in the next five years \citep{norris21}. 
Similarly, the Low-Frequency Array \citep[LOFAR;][]{vanharleem13} survey of the entire northern sky is expected to detect more than 10 million galaxies.
Other advanced radio telescopes include Murchison Widefield Array \citep[MWA;][]{wayth18}, MeerKAT \citep{jonas16} and the Karl G. Jansky Very Large Array \citep[JVLA][]{perley11}.
With the advent of the Square Kilometre Array (SKA\footnote{https://www.skatelescope.org/the-ska-project/}) radio telescope, which is expected to become operational in the coming years, the number of galaxy detections is expected to increase further, potentially reaching hundreds of millions.
Such an enormous dataset will significantly impact our understanding of the physics of galaxy evolution. 
This is set to significantly impact our understanding of the evolution of the universe.
However, to optimize the outcomes of these surveys, there is a need to innovate and develop new technologies for handling large datasets.


Radio galaxies are characterized by giant radio emission regions that extend well beyond their host galaxy structure at visible and infrared wavelengths. 
While most radio galaxies typically appear as simple, compact circular sources, increasing the sensitivity of radio telescopes results in the detection of more radio galaxies with complex extended structures. 
These structures typically consist of multiple components with distinct peak radio emissions. 
Based on the Fanaroff-Riley classifications \citep{fanaroff74}, radio galaxies exhibit two distinct morphologies, namely, Fanaroff-Riley Class I (FR-I) and Class II (FR-II) Active Galactic Nuclei (AGN) sources. FR-I sources have dark edges, causing the collimated jets from the central black holes at the center of the host galaxy to exhibit lower luminosity at the edges. FR-II sources, on the other hand, have brighter edges compared to the central host galaxies and sometimes lack a visible connection to the host galaxy.
Constructing catalogues of radio galaxies necessitates the grouping of associated components within extended radio galaxies and the identification of their corresponding infrared host galaxies.
Grouping these components is essential for estimating the real number density and overall integrated flux of radio galaxies. This process is crucial for modelling galaxy evolution and the expansion of the Universe. Failure to properly group associated components can lead to the misestimation of number density and total flux, resulting in inaccurate models.

This highlights the critical need for developing computer vision methods, to reduce the dependency on visual inspections to group associated radio source components and identify their infrared host galaxies.
Computer vision tasks are typically dependent on the available data. 
In supervised learning tasks, the model undergoes training by utilizing pairs of images and labels, where these labels provide precise and comprehensive information necessary for the model to make specific predictions.
Recently, these machine learning (ML) methods have been applied in the morphological classification and identification of radio sources, as demonstrated in studies such as \citep[e.g.][]{lukic18, alger18,wu19,bowles20,viera21,becker21,brand23}.
Self-supervised learning utilizes unsupervised techniques to train models based on the inherent data structure, removing the need for explicit annotations. This approach has proven effective in uncovering novel galaxy types in radio surveys, as demonstrated in studies such as \citep[e.g.][]{galvin20, mostert21, gupta22}.
In contrast, semi-supervised learning integrates both labelled and unlabelled data throughout the training process, as exemplified in \citet{slijepcevic22}.
In the realm of weakly supervised learning, indirect labels are utilized for the entire training dataset, serving as a supervisory signal. This particular approach has found utility in the classification and detection of extended radio galaxies \citep[][]{gupta2023a}.

As supervised learning relies on exact ground truth labels for training, this yields results that are more reliable in contrast to approaches that lack or possess limited supervisory signals during training.
However, to train and test such supervised algorithms, a large and diverse dataset of labelled radio galaxy images is necessary. Unfortunately, such a dataset is not currently available for next-generation radio surveys, which poses a significant challenge to developing automated methods for grouping multiple components of radio galaxies and identifying host galaxies. 
Therefore, there is a dire need for the development of a comprehensive dataset of radio galaxy images, along with accurate and consistent labels, to enable the development of effective machine learning algorithms for the automated grouping and identification of associated components of extended radio galaxies and their hosts.

This paper introduces RadioGalaxyNET, which includes a novel dataset and computer vision algorithms designed to tackle the challenge of associating radio galaxy components and identifying their hosts. 
Notably, our dataset is the first to be entirely curated by professional astronomers through multiple visual inspections and includes multimodal images of radio and infrared sky, as well as pixel-level labels on associated components of extended radio galaxies and their infrared hosts.
RadioGalaxyNET dataset has been structured in the COCO dataset format \citep{lin14M}, allowing for straightforward comparison studies of various object detection strategies for the machine learning community.
It includes 2,800 images with three channels each, incorporating two radio sky channels, one corresponding to the infrared sky, and a total of 4,155 annotated instances. The annotations encompass galaxy class information, bounding box details for capturing associated components of each radio galaxy, radio galaxy segmentation masks, and the host galaxy positions in infrared images.

To summarize, our work contributes to the following aspects:
\begin{itemize}
\setlength\itemsep{0.5em}
\item We introduce the first dataset entirely curated by professional astronomers that includes state-of-the-art images from the highly sensitive ASKAP telescope and instance-level annotations for extended radio galaxies.
\item As a novel addition, our dataset also includes corresponding images of the infrared sky, along with the positional information of the host galaxies.
\item We train and evaluate our dataset on seven cutting-edge object detection algorithms to demonstrate the challenge of detecting and grouping components of radio galaxies. Additionally, we propose a novel method to detect the positions of infrared host galaxies simultaneously.
\end{itemize}

The paper is organized as follows. In Section~\ref{SEC:dataset}, we provide details on the radio and infrared images, as well as the annotations utilized for training and evaluation. This section also offers a summary of the related work in this domain. Section~\ref{SEC:Methods} is dedicated to explaining the novel computer vision algorithms employed. Section~\ref{SEC:Training} provides comprehensive information about the network training process and the evaluation metric. In Section~\ref{SEC:Results}, we present the outcomes derived from the trained networks. Our findings are summarized in Section~\ref{SEC:Conclusions}, where we also outline directions for future research.

\begin{figure*}
\centering
\includegraphics[width=17cm, scale=0.5]{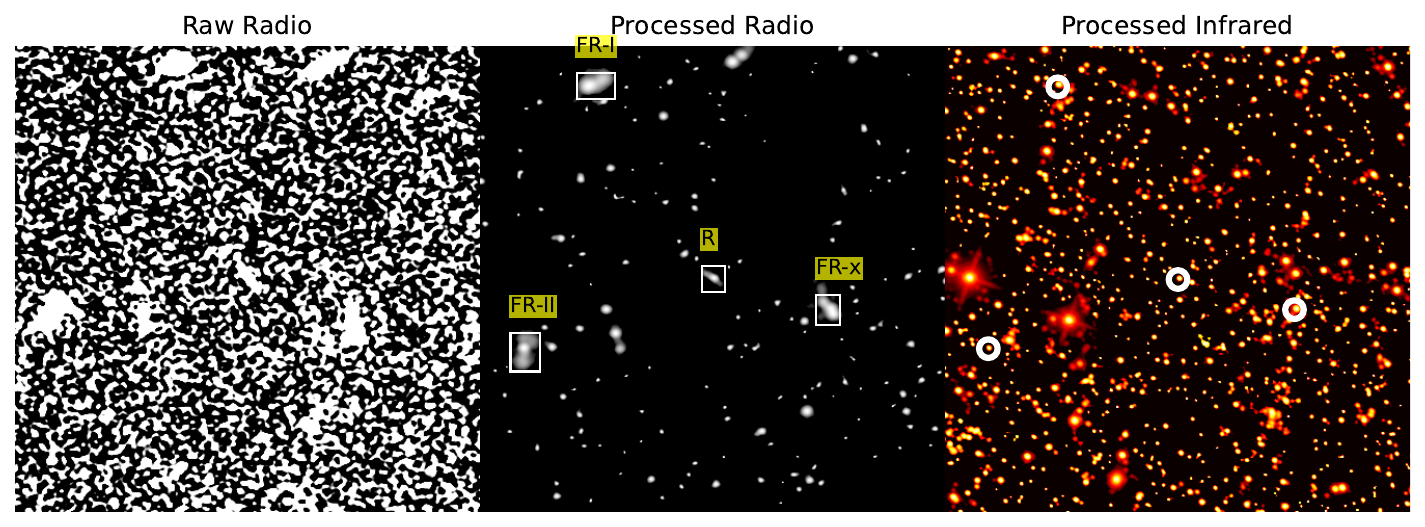}
\includegraphics[trim=0cm 0cm 0cm 1.15cm, width=17cm, scale=0.5]{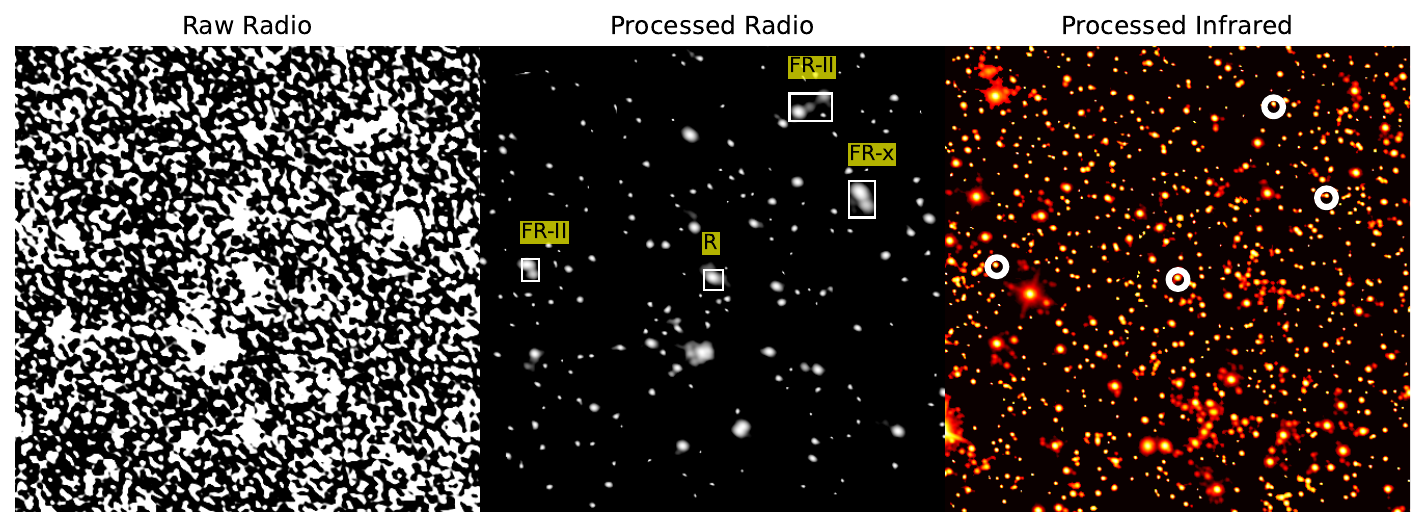}
\includegraphics[trim=0cm 0cm 0cm 1.15cm, width=17cm, scale=0.5]{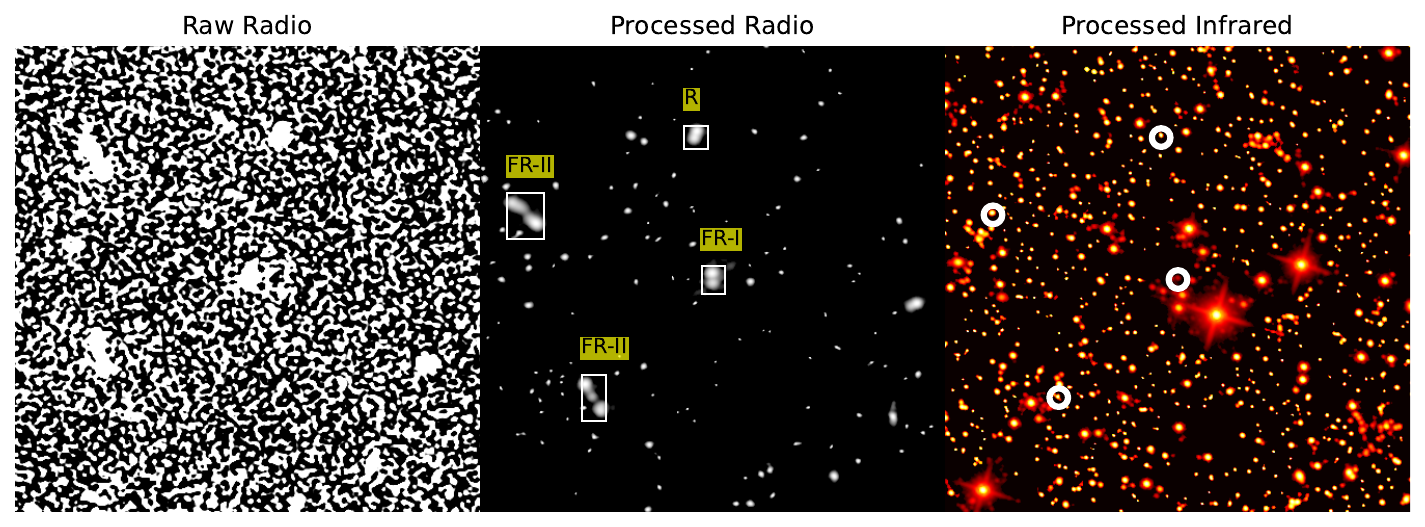}
\includegraphics[trim=0cm 0cm 0cm 1.15cm, width=17cm, scale=0.5]{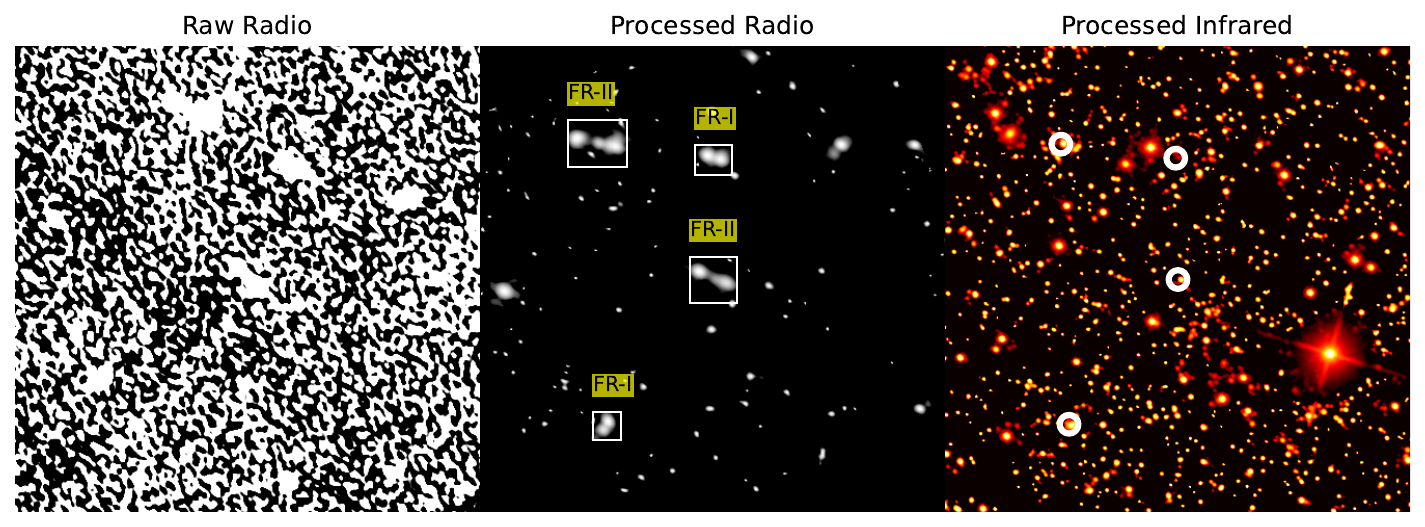}
\caption{Raw radio (left column), processed radio (middle column) and processed infrared (right column) images with the frame size of $450\times450$ pixels ($0.25^{\circ}\times 0.25^{\circ}$). The processed radio images highlight the categories of extended radio galaxies, and the bounding boxes denote their total radio extent encompassing all of its components. The infrared images show host galaxies inside the circles.} 
\label{FIG:RadioGalaxies15arcm}
\end{figure*}

\section{RadioGalaxyNET: Dataset}
\label{SEC:dataset}
\subsection{Radio and Infrared Images}
The RadioGalaxyNET dataset contains radio images derived from observations with the ASKAP radio telescope. 
ASKAP is located at Inyarrimanha Ilgari Bundara, the Murchison Radio-astronomy Observatory (MRO). Utilizing advanced technologies like the phased array feed (PAF) \citep[PAF][]{hay06}, ASKAP can efficiently scan vast sections of the sky, significantly enhancing survey capabilities. Comprising 36 antennas, ASKAP has baselines reaching up to 6.4 km, with 30 antennas concentrated within a 2.3 km radius \citep{hotan21}. ASKAP is currently involved in various surveys, each designed for specific scientific objectives. One notable survey is the EMU survey, which aims to identify approximately 40 million galaxies over its five-year operational period. In 2019, the EMU pilot survey was conducted to develop technologies in preparation for the main survey.


The EMU-PS survey encompasses a sky area of 270 deg$^2$, ranging from $301^{\circ}$ to $336^{\circ}$ in Right Ascension and from $-63^{\circ}$ to $-48^{\circ}$ in Declination. 
Comprising ten overlapping tiles, each tile underwent a total integration time of approximately 10 hours, resulting in an RMS noise of $25-35~\mu$Jy/beam. 
The survey operated within a frequency range of 800 to 1088 MHz, centered at 944 MHz (with a wavelength of 0.37 to 0.28m, centered at 0.32m). 
The telescope's visibility data, recording electric fields as a function of time and frequency, was processed through the ASKAPsoft pipeline \citep[][]{whiting17} to generate images for the EMU-PS.

The identification of the 2,800 extended radio galaxies in the EMU-PS image involved a meticulous manual process conducted in three distinct stages, as detailed by \citet{yew22prep}. 
While a comprehensive overview of the source identification methodology is available in their work, we provide a brief summary here. Initially, a quick assessment of the EMU-PS radio image was conducted to identify clearly visible extended sources. 
Subsequently, a thorough visual scan covered the entire radio image systematically, involving the classification of sources and marking the extent of diffuse radio emission for each. 
In the third and final stage, an exhaustive scan aimed to uncover additional sources possibly overlooked in previous stages. 
Upon completion of the identification process, the centroid position and approximate size of each source were documented. 
It is important to note that, despite the three rounds of manual identification, locating all extended sources remains challenging, particularly for faint ones, given the substantial scientific resources required for such an exhaustive search.
Following this, cutouts of size $0.25^{\circ}\times 0.25^{\circ}$ were generated at the centroid positions of these galaxies, resulting in $450\times 450$ pixel images with a pixel size of 2 arcseconds. 
The left column of Figure~\ref{FIG:RadioGalaxies15arcm} presents examples of noisy raw images.
Following the procedure outlined by \cite{gupta2023a}, we preprocess the raw images. 
Examples of processed radio images are illustrated in the middle column of Figure~\ref{FIG:RadioGalaxies15arcm}. 

At the same sky locations of radio images, we obtain AllWISE \citep[][]{cutri13} infrared images from the Wide-field Infrared Survey Explorer's \citep[WISE;][]{wright10} W1 band that correspond to 3.4 $\mu$m wavelength.
Similar to radio images, infrared images are processed to reduce noise in the same way. 
However, the noise in the infrared images is estimated as Median $+~3\times$MAD due to the non-Gaussian nature of the noise and the saturation of sources in the images.
The right column of Figure~\ref{FIG:RadioGalaxies15arcm} presents examples of processed infrared images, while raw infrared images are not displayed for brevity.

Finally, we create 3-channel RGB images by combining the processed radio and infrared images.
In this process, the B and G channels represent radio channels. 
The original 32-bit FITS image is initially converted to 16-bit, and its 8-16 bit and 0-8 bit information are assigned to the B and G channels, respectively. 
Similarly, the infrared FITS image undergoes a conversion to 16-bit, and its 8-16 bit information is incorporated into the R channel.

\begin{figure*}
\centering
\includegraphics[width=17.5cm, scale=0.5]{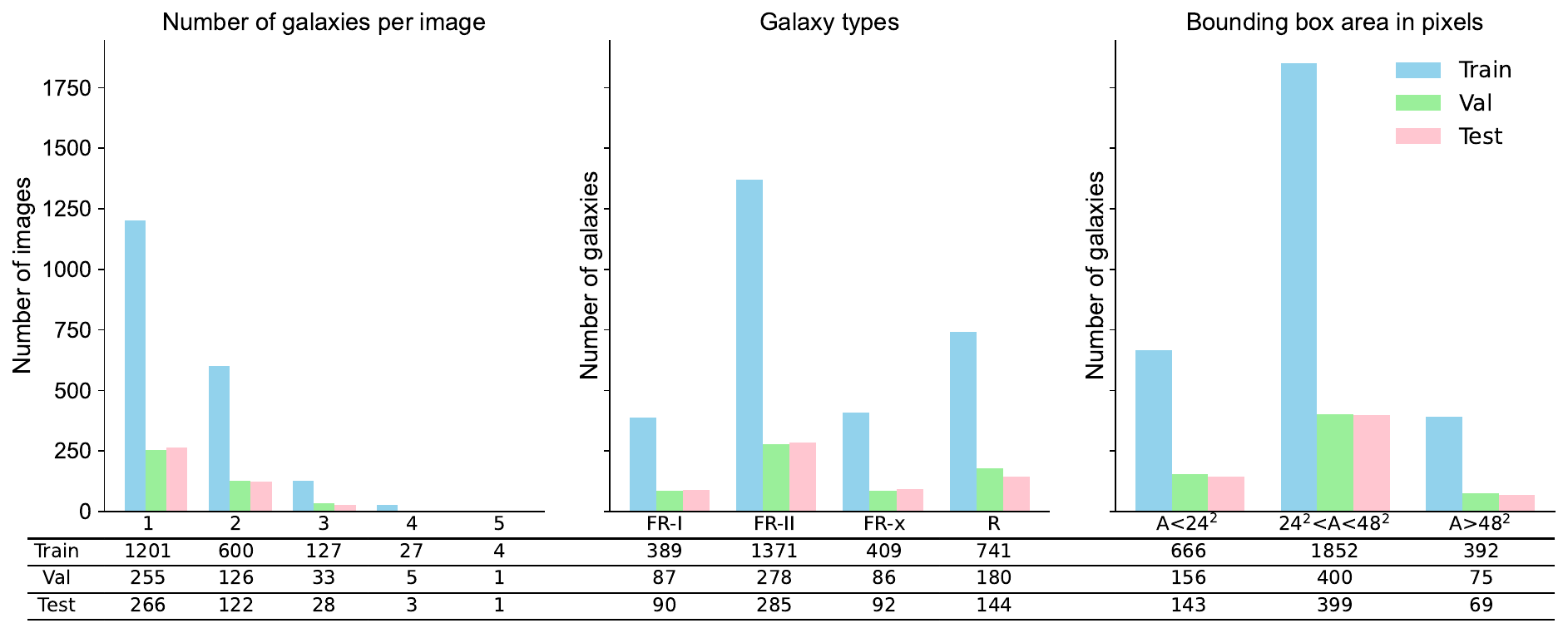}
\caption{The dataset split distributions of the RadioGalaxyNET. Shown are the distributions of extended radio galaxies in one frame (left), their categories (middle) and the occupied area per radio galaxy (right). The tables presented below the figures display the precise counts of galaxy instances within the training, validation, and test sets.}
\label{FIG:Statistics}
\end{figure*}

\subsection{Annotations}

The RadioGalaxyNET dataset has the following four sets of labels
\begin{itemize}
    \item the infrared host galaxy positions.
    \item the extended radio galaxy classes,
    \item the bounding boxes enclosing all components of each radio galaxy, and
    \item the segmentation masks for radio galaxies.
\end{itemize}

The labelling process details are discussed in \cite{yew22prep} and \cite{gupta2023a}. 
In summary, we visually identified host galaxies in the infrared images, with the most likely hosts situated near or over the central peaks of radio emission. 
Only sources with identified host galaxies in the infrared image were included in the dataset. Radio sources were classified based on criteria from \cite{fanaroff74}, distinguishing between FR-I and FR-II sources by considering the distance between the peak radio emission of all components and the total extent of the source. 
Additionally, some sources were classified as FR-x due to uncertainties in determining peak flux positions and total extent, as explained in \citep{gupta2023a}. 
Another category, denoted as R, includes resolved sources exhibiting extended emissions without clear peaks other than in the central part.
The bounding boxes for each radio source, measured with the CARTA visualization package \citep{comrie21}, served as the basis for obtaining segmentation masks. These masks were then created for all pixels within the bounding box, where the flux exceeded 3$\sigma$.
Annotations in the RadioGalaxyNET dataset adhere to the COCO dataset format \citep{lin14M}, facilitating straightforward comparisons of object detection methods. 
Radio annotations for each galaxy are stored as `categories', `bbox', and `segmentation', while the positions of the infrared hosts are stored as `keypoints'.

\subsection{Data Statistics and Possible Tasks}
\label{SEC:Datastats}
The RadioGalaxyNET dataset encompasses 2,800 3-channel images featuring two radio sky channels and one corresponding infrared sky channel. 
Both noisy and processed radio images are included in the dataset. 
It comprises 2,800 extended radio galaxies, resulting in a total of 4,155 instances of these galaxies due to their proximity in the sky and their appearance in multiple images. 
The radio galaxies in the dataset are categorized as 13\% FR-I, 48\% FR-II, 14\% FR-x, and 25\% R sources.
The dataset is divided into three sets: training, validation, and test sets, with a split ratio of $0.7:0.15:0.15$, respectively. 
This split ratio is widely used in machine learning and represents a balanced approach that achieves a reasonable trade-off between having sufficient data for training and fine-tuning the model, while also ensuring that the model is evaluated on a suitably large, independent test set.
The statistics for the number of objects in one frame, categories of extended radio galaxies, and the occupied area of labelled objects are shown in Figure~\ref{FIG:Statistics}. 
The distribution of target galaxies is approximately matched between the training, validation, and test sets. 
Small extended radio galaxies (area $< 48^2$ in pixels) make up the largest proportion of the dataset.

Although the primary goal of this study is to detect radio galaxies, the RadioGalaxyNET dataset's unique multimodality can be leveraged for the simultaneous identification of infrared host galaxies. 
Additionally, it could be insightful to explore approaches for grouping multiple components of galaxies. 
Instead of component association using given box boundaries, the direct segmentation of related galaxy components could present a greater challenge for models. 
This could also be linked to instance segmentation approaches, where different input modalities could provide an interesting challenge.
The dataset's unlabelled extended radio galaxies in a given frame make it suitable for experimenting with semi-supervised and contrastive learning methods.
Furthermore, RadioGalaxyNET dataset could be used to validate signal enhancement in previous radio sky surveys. 
By downweighing detections that resemble extended galaxies in the dataset images, the dataset could be used to discover new galaxies in the radio sky. 
Finally, the modalities in the dataset could be used to model extended radio galaxy emissions from infrared hosts and vice versa.

In this paper, we only focused on extended radio galaxies for the current dataset and have not included labels for several compact sources in radio images.
Few extended radio galaxies, such as Odd Radio Circles \citep[][]{norris21c}, some Giant Radio Galaxies and other peculiar galaxies \citep[][]{gupta22}, are also not included in the dataset. 
Future efforts should focus on including a balanced representation of these sources in the dataset.

\subsection{Related Work}
\label{SEC:Related}
\subsubsection{Existing Radio Galaxy Dataset}
The availability of datasets for detecting and classifying radio galaxies is limited. 
The MiraBest Batched Dataset is a labelled dataset of FR galaxies extracted from \cite{miraghaei17}. 
It contains FR-I, FR-II, and hybrid extended radio sources, with any nonstandard morphology reported. 
The dataset comprises a total of 1,256 images obtained from the VLA FIRST sky survey \citep{becker95}, with all images measuring $150 \times 150$ pixels, with one-pixel corresponding to an angular size of 1.8 arcseconds. 
The dataset is divided into seven training batches and one test batch, with each batch consisting of approximately 157 images, and the number of objects in each class is distributed relatively evenly across batches. 
The test batch contains at least one example of each class. 
It is important to note that the MiraBest dataset only provides information about the classes of radio galaxies without annotations for bounding boxes, segmentation masks, or infrared hosts, making it challenging to apply to multi-component association problems.

Another dataset is sourced from the Radio Galaxy Zoo (RGZ) citizen science initiative, as described by \citep[][]{wu19}. 
Over 12,000 volunteers visually identified radio galaxies, resulting in a dataset of 6,536 extended radio galaxies with a user-weighted consensus level (CL) of at least 0.6. 
The CL reflects the level of agreement among citizens regarding the classification. 
The images in this dataset were obtained from the FIRST radio survey, and the annotations include the position of galaxies in the sky and the bounding boxes derived from web portal clicks by volunteers. 
These boxes are saved as the approximate angular size in the RGZ dataset. However, the dataset does not provide information on the segmentation masks and infrared host positions. 
The RMS noise in the FIRST radio images is approximately ten times larger than that in the EMU-PS, resulting in a lower density of galaxies in the same area.
This high galaxy density makes RadioGalaxyNET dataset images derived from EMU-PS challenging for object detection.
Table~\ref{TAB:DataComparison} shows a comparison between the existing and our new dataset.

It is important to emphasize that the previous dataset possesses radio image noise levels that are an order of magnitude higher than our dataset. This higher noise level renders low-intensity extended structures essentially invisible. As a result, these previous datasets are not optimal for training networks aimed at detecting radio galaxies in the next generation of radio telescopes. 
\begin{table*}
  \caption{Datasets currently available for the machine learning tasks of classification and object detection involving radio galaxies. The annotations C, B, S, and K are categories, bounding boxes, segmentation and keypoint labels, respectively.
  Section~\ref{SEC:dataset} provides a detailed description of the annotations for both our dataset and the existing dataset.}
  \label{sample-table}
  \centering
  \begin{tabular}{lccccc}
    \toprule \\
    Name                    & \#Extended & Annotation      & Radio Image & Domain  & Image Size\\
                            & Galaxies  & type        & noise ($\mu$Jy/b) & experts      & (pixels)\\
    \midrule
    MiraBest                & 1,256     & C           & $\sim140$  & \cmark & $150\times150$\\
    Citizen Science RGZ     & 6,536     & C, B        & $\sim140$  & \xmark & $132\times132$\\
    RadioGalaxyNET (ours)   & 2,800      & C, B, S, K  & $\sim30$   & \cmark & $450\times450$\\
    \bottomrule
  \end{tabular}
  \label{TAB:DataComparison}
\end{table*}

\subsubsection{Existing Machine Learning Applications}
\label{SEC:Related2}
In recent years, radio galaxy detection in images has attracted increasing attention, particularly focusing on deep learning algorithms. 
These algorithms are commonly divided into two categories: region-proposal-based methods and classification-based methods.
\cite{wu19} used Faster Region-based Convolutional Neural Network \citep[Faster-RCNN;][]{Shaoqing15} based method to locate and classify radio galaxies in the RGZ dataset.
\cite{lao21} used a combination of residual neural network \citep[ResNet;][]{he16CVPR} and Feature Pyramid Network \citep[FPN;][]{lin17CVPR} to detect and classify FR galaxies.
\cite{scaife21} classified FR galaxies in the MiraBest dataset using Group equivariant Convolutional Neural Networks \citep[G-CNNs;][]{cohen16ICML}.
\cite{slijepcevic22} used FixMatch \citep{sohn2020fixmatch} to classify both MiraBest and RGZ radio galaxies.
\cite{zhang22} used You Only Look Once method \citep[YOLOv5;][]{redmon16CVPR} to classify radio galaxies.

\begin{figure*}
\centering
\includegraphics[width=18.cm, scale=0.5]{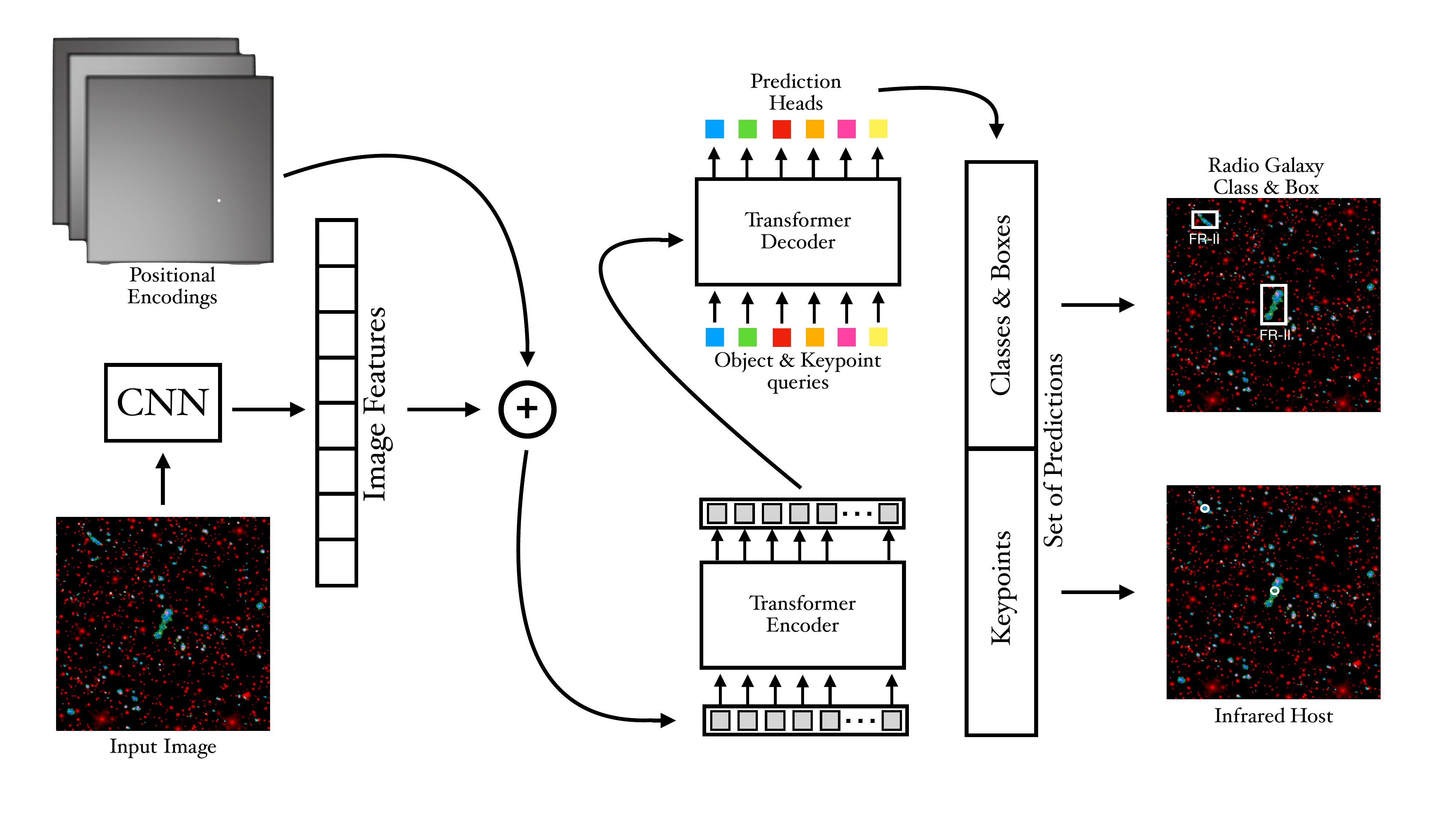}
\caption{An overview of the multimodal modelling strategy introduced in this study. In the context of the Gal-DETR model (refer to Section~\ref{SEC:GalDETR}), we introduce a keypoint estimation module within the transformer encoder-decoder framework. This enables the simultaneous detection of categories and bounding boxes for radio galaxies, and the positions of infrared hosts. A similar multimodal strategy is introduced for Gal-Deformable DETR and Gal-DINO (as detailed in Sections~\ref{SEC:GaldDETR} and \ref{SEC:GalDINO}).} 
\label{FIG:NET_Overview}
\end{figure*}

\section{RadioGalaxyNET: Novel Computer Vision Methods}
\label{SEC:Methods}
While the machine learning techniques employed in prior studies (refer to Section~\ref{SEC:Related2}) have their own strengths. 
Recent advancements in computer vision have introduced several more effective algorithms that outperform the previously utilized methods in real-world scenarios. 
In addition, none of the previously employed techniques possess the ability to concurrently detect both multiple-component radio galaxies and their corresponding infrared hosts.
In this section, we describe these state-of-the-art object detection approaches and explain our novel enhancements to these methods, which allow for the simultaneous detection of both radio galaxies and infrared hosts.

\subsection{Gal-DETR}
\label{SEC:GalDETR}
In this paper, we introduce Gal-DETR, multimodal model for computer vision. The Gal-DETR model consists of two primary components: the DEtection TRansformers (DETR), as described in \citet[][]{carion2020end} that is trained for the class and bounding box prediction for radio galaxies, and our novel Keypoint detection module, trained concurrently to detect infrared hosts.
Note that the existing multimodal methods are tailored to specific tasks. Here we have radio images where galaxies appear larger due to extended emission, while in infrared images, the host galaxies look like point objects (as depicted in columns 2 and 3 of Figure~\ref{FIG:RadioGalaxies15arcm}). To the best of our knowledge, there are no specific models that deal with objects that look completely different in two image modalities. As a result, we introduce our own approach to modelling, illustrated in Figure~\ref{FIG:NET_Overview}.

\subsubsection{Class and Bounding Box Detection for Radio Galaxies}
DETR model leverages the Transformer architecture \citep[][]{vaswani2017attention}, initially developed for natural language processing, to tackle the complex task of object detection in images following \citep[][]{dosovitskiy2020image}. 
Unlike traditional methods that rely on region proposal networks \citep[e.g. Faster RCNN;][]{ren2015faster}, DETR introduces end-to-end object detection using Transformers.
The model begins with a convolutional neural network (CNN) backbone, such as ResNet, to process the input image and extract relevant feature maps.
ResNet-50 is a convolutional neural network architecture designed to address the challenge of vanishing gradients in very deep neural networks. It incorporates 50 layers and utilizes skip connections, also referred to as residual connections, to enhance gradient flow during training \citep{he15d}.
To incorporate spatial information into the Transformer architecture, DETR introduces positional encodings. These encodings are added to the feature maps, providing the model with information about the relative positions of objects within the image.
The feature maps, enhanced with positional encodings, are then passed through a Transformer encoder. This component enables the model to simultaneously process the entire image, capturing contextual relationships between various regions.
DETR introduces learned object queries, similar to class-specific anchor boxes in traditional detectors \citep[e.g.][]{tian2019fcos}. 
Note that the anchor boxes are prior boxes defined as a set of pre-determined, fixed-size boxes of different scales and aspect ratios. These anchor boxes are chosen based on prior knowledge of the expected object sizes and shapes within the dataset. 
However, unlike fixed anchor boxes, the object queries in DETR are learned during training through the transformer decoder and represent the classes of objects the model aims to detect.
The final output from the decoder has two distinct output heads: one for class prediction, which estimates the class probabilities for each object query, and another for bounding box prediction, which calculates the coordinates (x, y, width, height) of each object query's bounding box (see Figure~\ref{FIG:NET_Overview}).

DETR employs a Hungarian loss\footnote{https://github.com/mmgalushka/hungarian-loss} function to establish associations between predicted bounding boxes and ground-truth boxes. This loss enforces a one-to-one mapping between predicted and ground-truth objects, ensuring the model's permutation invariance.
The overall loss function for DETR combines the classification loss (cross-entropy) for class predictions and the box loss (smooth L1 loss) for bounding box predictions,
\BE
\mathcal{L}_{\text{DETR}} = \mathcal{L}_{\text{class}} +  \mathcal{L}_{\text{box}},
\label{EQ:LossDETR}
\EE
where $\mathcal{L}_{\text{class}}$ is the cross-entropy loss for class predictions and $\mathcal{L}_{\text{box}}$ is the smooth L1 loss for bounding box predictions.
In the context of object detection, the L1 loss is often used for bounding box regression. It calculates the absolute difference between the predicted bounding box coordinates (e.g., x, y, width, height) and the ground truth bounding box coordinates. This loss penalizes the model for deviations between the predicted and true bounding box values, with larger differences leading to higher loss values.

\subsubsection{Keypoint Detection for Infrared Hosts}
In addition to the object detection method that employs bounding boxes to detect extended radio galaxies, the integration of keypoint detection techniques \citep[e.g.][]{Simon_2017_CVPR} can offer a complementary approach for identifying the corresponding infrared hosts. 
Keypoints, representing specific landmarks or distinctive features within the radio and infrared images, provide valuable spatial information that aids in precisely determining the location of the host galaxy. 
Unlike bounding boxes that enclose the entire source, keypoints allow for a more fine-grained localization, providing a more accurate representation of the host's position within the radio source. 
This precise localization is especially valuable when the radio emission exhibits complex morphologies.

In the present work, we extended the capabilities of DETR by incorporating keypoint detection.
We refer to this enhanced version as Gal-DETR.
In Gal-DETR, we integrated keypoint detection to complement the existing object detection capabilities of the original DETR algorithm (see Figure~\ref{FIG:NET_Overview}). 
By incorporating keypoint detection, Gal-DETR gains the ability to identify salient keypoints within the objects, providing additional spatial information and fine-grained details.
The keypoint detection module in Gal-DETR leverages the transformer-based architecture of DETR to capture global context and local details. 
By attending to the keypoint embeddings and utilizing self-attention mechanisms, Gal-DETR localizes and associates keypoints for the infrared host galaxies.
The overall loss function for Gal-DETR combines the DETR loss for class and bounding box predictions, in addition to the keypoint detection loss as
\BE
\mathcal{L}_{\text{Gal-DETR}} = \mathcal{L}_{\text{DETR}} +  \mathcal{L}_{\text{keypoint}}
\label{EQ:LossGalDETR}
\EE
where $\mathcal{L}_{\text{keypoint}}$ is the L1 loss for keypoint detection to detect infrared hosts. In the context of keypoint detection, the L1 loss calculates the absolute difference between the predicted keypoint coordinates (e.g., x, y position of host) and the ground truth keypoint coordinates.

\subsection{Gal-Deformable DETR}
\label{SEC:GaldDETR}
The second multimodal model we introduce in this paper is Gal-Deformable DETR. Similar to Gal-DETR, the Gal-Deformable DETR model consists of two primary components: the Deformable DEtection TRansformers (Deformable DETR), as described in \citet[][]{zhu2021deformable} that is trained for the class and bounding box prediction for radio galaxies, and our novel Keypoint detection module, trained concurrently to detect infrared hosts.

Deformable DETR builds upon the foundation of DETR and introduces several key enhancements.
Deformable attention mechanisms allow the model to adaptively adjust the spatial sampling locations for better feature extraction, especially for objects with complex shapes or poses.
Unlike DETR, which relies solely on learned positional encodings, Deformable DETR adds spatial positional encodings to the feature maps. This addition helps the model better capture the spatial relationships between objects.

In this study, we have expanded the functionalities of Deformable DETR by introducing keypoint detection, resulting in an enhanced version known as Gal-Deformable DETR. 
Much like the Gal-DETR model, Gal-Deformable DETR incorporates keypoint detection as a complementary component to the existing object detection capabilities inherited from the original Deformable DETR algorithm. 
Specifically, we have integrated keypoint detection in conjunction with the deformable convolutional layers originally introduced in the Deformable DETR framework. 
Similar to Equation~\ref{EQ:LossGalDETR}, the L1 loss for keypoint detection is combined with the class and bounding box loss during the training of the Gal-Deformable DETR model.

\subsection{Gal-DINO}
\label{SEC:GalDINO}
The third multimodal model we introduce in this paper is Gal-DINO. Following Gal-DETR and Gal-Deformable DETR models, Gal-DINO also consists of two primary components: the DETR with Improved deNoising anchOr boxes (DINO), as described in \citet[][]{zhang2022dino} that is trained for the class and bounding box prediction for radio galaxies, and our novel Keypoint detection module, trained simultaneously to detect infrared hosts.

DINO extends the DETR model with improved denoising anchor boxes, introducing several key enhancements.
DINO improves anchor boxes, which are pre-defined boxes used for object detection. It introduces better strategies for selecting and placing these anchor boxes, enhancing the model's ability to detect objects of different sizes and aspect ratios.
DINO introduces an improved mechanism for matching anchor boxes to ground truth objects during training, making the model more accurate in localization and classification.
DINO employs adaptive convolutional features, allowing the model to focus on informative regions of the image, thus improving both efficiency and accuracy.

Much like our approach with Gal-DETR, we integrated keypoint detection into the DINO algorithm, which already has improved de-noising anchor boxes. 
This enhanced version, featuring keypoint detection, is denoted as Gal-DINO. 
By reducing the impact of noise and outliers, Gal-DINO produces more robust and precise bounding box predictions, resulting in better localization of the extended radio sources and their associated infrared hosts within these enhanced bounding boxes.
In a manner similar to the approach outlined in Equation~\ref{EQ:LossGalDETR}, the Gal-DINO model incorporates the L1 loss for keypoint detection combined with the class and bounding box loss during the training process.

\subsection{Gal-SIOD}
\label{SEC:GalSIOD}
Detection of objects in situations with imperfect data has recently become a focal point. 
Weakly supervised object detection (WSOD) encounters notable challenges in localization due to the lack of direct annotations \citep[e.g.][]{gupta2023a}.  
\citet{li2022siod} introduced an approach known as Single Instance Annotated Object Detection (SIOD). 
We refer the reader to their paper for detailed information about the SIOD method.
Briefly, SIOD requires just one instance annotation for each category present in an image.
SIOD offers a more robust and informative source of prior knowledge for detecting the remaining unlabelled instances. 
This approach strikes a balance between annotation cost and performance, providing a valuable solution to object detection challenges under imperfect data conditions.

This holds significance in the context of our current study on radio galaxy detection.
As detailed in Section~\ref{SEC:dataset}, even after conducting three rounds of manual search of EMU-PS, not all extended sources could be precisely located.
This arises from the fact that visual inspections entail an exhaustive search for faint extended radio galaxies, a process demanding a substantial allocation of scientific resources.
This implies that certain extended and peculiar sources within the same image are not annotated in the dataset.
Utilizing SIOD for radio galaxy detection offers us the possibility to detect these unlabelled extended radio galaxies within a given image.

Note that we have not incorporated keypoint detection into the radio galaxy detection process using SIOD. 
This decision stems from our primary objective, which is to assess whether we can detect additional radio galaxies not labelled in the dataset using this method. 
Future work should integrate keypoint detection into SIOD to enable the simultaneous detection of infrared hosts alongside radio galaxy detections. 
While our current implementation directly employs SIOD for our task, we have named our implementation as Gal-SIOD for naming consistency.

\subsection{Gal-SIOD-DMiner}
\label{SEC:GalSIODD}
SIOD-DMiner \citep[][]{li2022siod} improves upon SIOD by introducing the Dual-Mining (DMiner) framework. 
SIOD, while effective in certain scenarios, faces challenges when it comes to dealing with unlabelled regions. 
Directly assigning all unlabelled regions as background can adversely affect the training process and detector performance. 
SIOD-DMiner tackles this challenge and improves object detection efficacy by integrating a Similarity-based Pseudo-Label Generating module (SPLG). 
This module retrieves instances by assessing feature similarity between labelled reference instances and the remaining unlabelled area in the image.
This improves object detection over SIOD, which lacks this mechanism for leveraging unlabelled data effectively.

In addition, the SIOD-DMiner recognizes that relying solely on pseudo-labels generated by SPLG can be problematic. It can lead to confusion due to false pseudo-labels, especially when the model focuses on learning a hyperplane for discriminating each class from the others. To mitigate this issue, SIOD-DMiner introduces a Pixel-level Group Contrastive Learning module (PGCL). 
PGCL improves the model's ability to withstand inaccurate pseudo-labels, reducing reliance on potentially flawed annotations.
For similar considerations as mentioned earlier, we have not integrated keypoint detection into the radio galaxy detection process using SIOD-DMiner. 
Future work should integrate keypoint detection into SIOD-DMiner to enable the simultaneous detection of unlabelled infrared hosts.
Furthermore, we have employed the SIOD-DMiner from \citet{li2022siod} for our task but named our implementation Gal-SIOD-DMiner to maintain naming consistency.

\subsection{Gal-Faster RCNN}
\label{SEC:GalFRCNN}
Faster Region-based Convolutional Neural Network \citep[Faster RCNN;][]{ren2015faster} significantly improved the speed and accuracy of object detection compared to the proposal-based predecessor methods. 
Faster RCNN follows a two-stage approach to object detection, which distinguishes it from earlier methods like Fast RCNN \citep[][]{girshick2015fast}.
The first stage of Faster R-CNN is a Region Proposal Network (RPN), which efficiently generates region proposals (candidate bounding boxes) that are likely to contain objects of interest. The RPN operates on feature maps extracted from the input image and predicts regions with high scores.
In the second phase, the system fine-tunes and categorizes these region proposals. 
Region of Interest (RoI) pooling is employed to derive standardized feature maps of fixed dimensions from each region proposal, making them suitable for a standard CNN classifier.
Faster RCNN uses a CNN classifier to categorize the content within each RoI into predefined object classes (e.g., FR-I, FR-II etc.) and regress the bounding box coordinates for precise localization.
The entire network, including both the RPN and the classifier, is trained end-to-end. This means that the model learns to generate region proposals and classify objects simultaneously during training.
While Deformable DETR and DINO have demonstrated superior performance over Faster RCNN when applied to real-life images from the COCO dataset \citep[][]{zhu2021deformable, zhang2022dino}, we have opted to utilize Faster RCNN on our dataset for the sake of comparison. 
In line with this primary objective of comparing Faster RCNN with Transformer-based methods, we have refrained from incorporating Keypoint detection into Faster RCNN. 
We have named our implementation Gal-Faster RCNN to maintain naming consistency.

\begin{figure*}
\centering
\includegraphics[width=18.cm, scale=0.5]{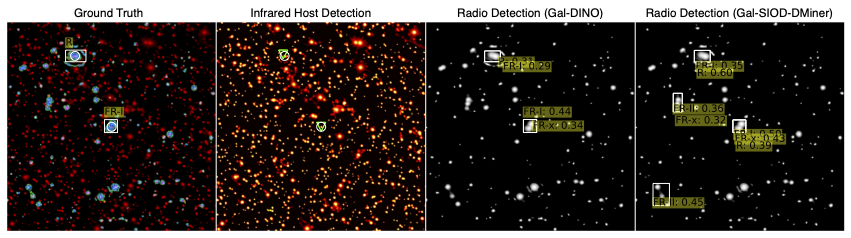}
\includegraphics[trim={0cm 0cm 0cm 1.55cm}, clip, width=18.cm, scale=0.5]{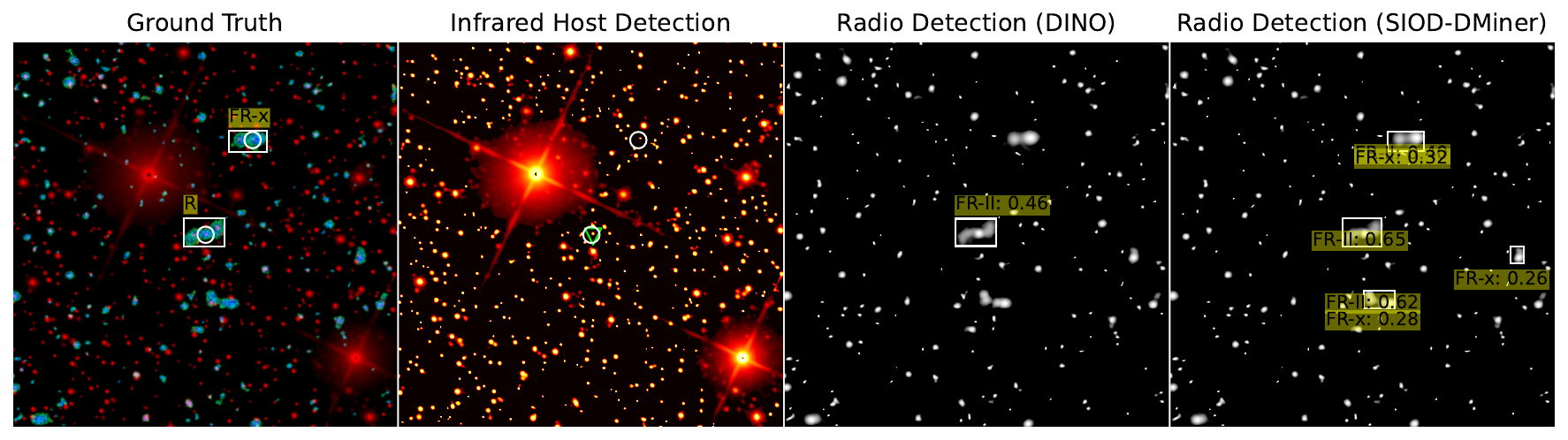}
\includegraphics[trim={0cm 0cm 0cm 1.55cm}, clip, width=18.cm, scale=0.5]{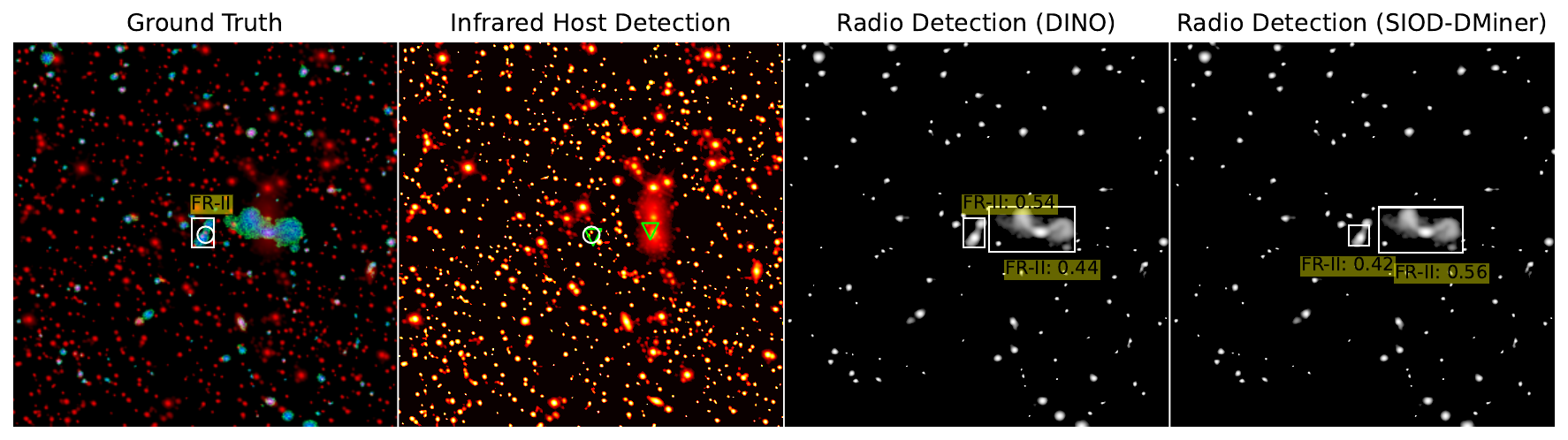}
\includegraphics[trim={0cm 0cm 0cm 1.55cm}, clip, width=18.cm, scale=0.5]{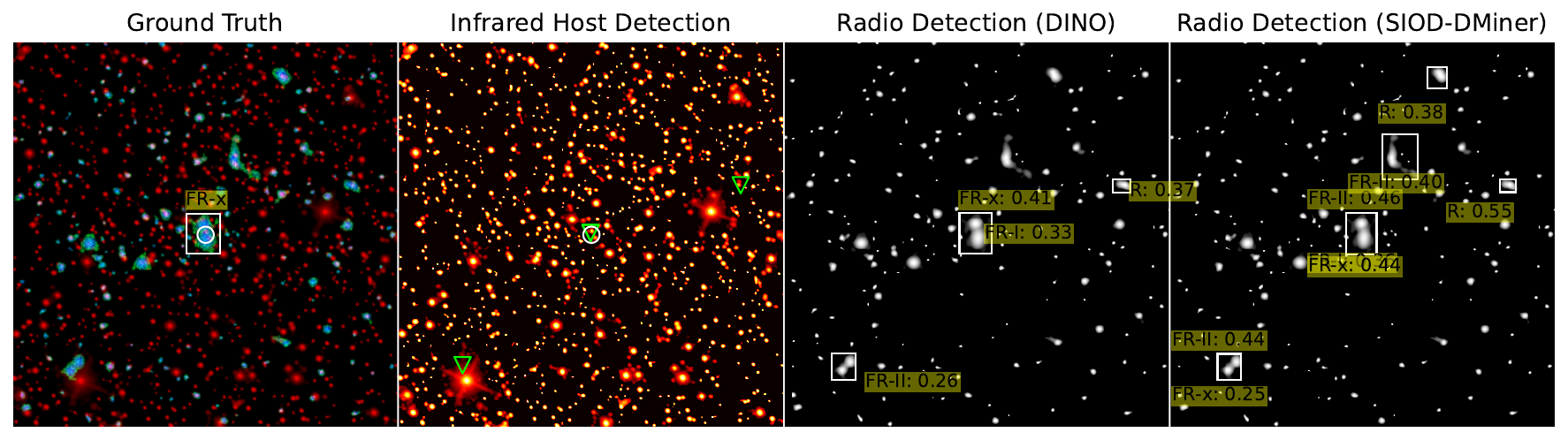}
\includegraphics[trim={0cm 0cm 0cm 1.55cm}, clip, width=18.cm, scale=0.5]{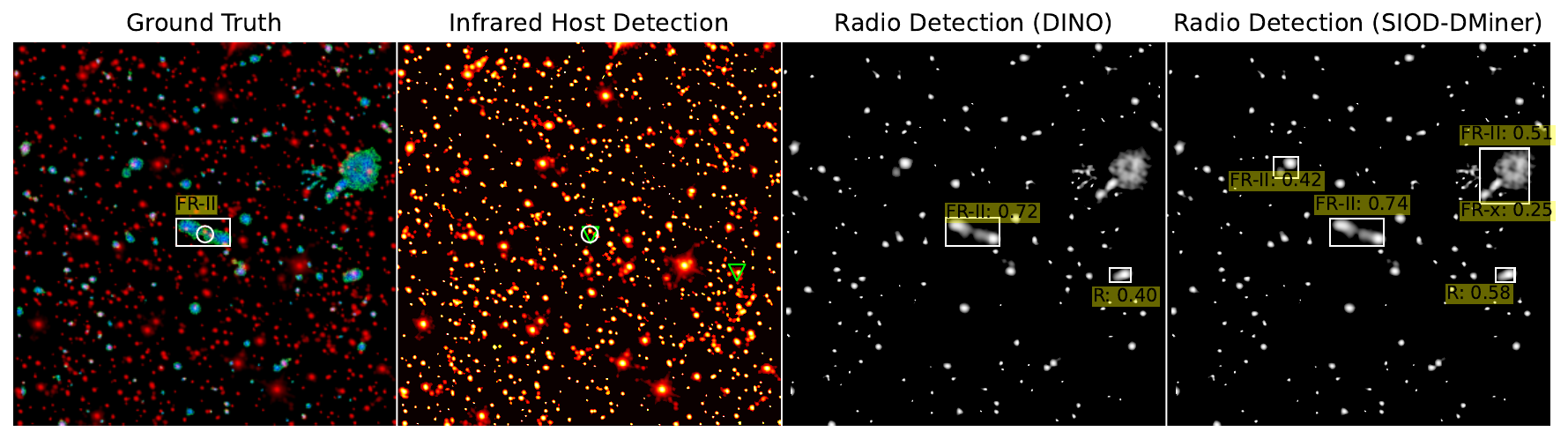}
\caption{Object detection results: Shown are the processed radio-radio-infrared images and ground truth annotations (first column), ground truth and Gal-DINO keypoint detections as circles and triangles over infrared images (second column), Gal-DINO (third column) and Gal-SIOD-DMiner (fourth column) class and bounding box predictions over radio images with a confidence threshold of 0.25. These models exhibit the capability to detect additional extended galaxies that lack ground truth annotations.} 
\label{FIG:Results}
\end{figure*}

\subsection{Gal-YOLOv8}
\label{SEC:GalYOLO}
You Only Look Once version 8 (YOLOv8) is an object detection model latest in the YOLO \citep[][]{redmon2016you} series of object detection models. 
YOLOv8 follows the one-stage detection paradigm, which means it directly predicts bounding boxes and class probabilities for objects in a single pass through the neural network. This makes it faster compared to two-stage detectors like Faster R-CNN.
YOLOv8 employs a CNN backbone architecture that is designed to capture features at multiple scales.
YOLOv8 uses a detection head consisting of convolutional layers to predict bounding boxes and class probabilities. It predicts class labels and confidence scores for each detected object.
Similar to initial YOLO versions, YOLOv8 uses anchor boxes to help predict the shape and location of objects. However, YOLOv8 has the option to automatically calculate anchor box sizes based on the dataset, which can simplify the training process.
In this study, we perform fine-tuning on a pre-trained YOLOv8 model using our dataset and designate our customized implementation as Gal-YOLOv8 to maintain naming consistency. 
It is worth noting that we have not integrated keypoint detection into YOLOv8 in this implementation. 
Future work should explore methods for detecting infrared hosts alongside radio galaxy detections within the YOLO class of models.

\begin{table*}
    \centering
    \caption{Bounding box and keypoint detection results on the test set of RadioGalaxyNET. From left to right, the columns display the multimodal models introduced in this study, the number of model parameters in millions, the number of training epochs, the average precision for IoU (or OKS) thresholds ranging from 0.50 to 0.95 (AP), a specific IoU (or OKS) threshold of 0.5 (AP$_{50}$), IoU (or OKS) threshold of 0.75 (AP$_{75}$), and the average precision for small-sized radio galaxies (AP$_{\rm S}$), medium-sized radio galaxies (AP$_{\rm M}$), and large-sized radio galaxies (AP$_{\rm L}$), categorized by areas less than $24^2$, between $24^2$ and $48^2$, and greater than $48^2$ pixels, respectively. Detailed information on the training and development of these models is provided in Section~\ref{SEC:Training}, while the models themselves are described in Section~\ref{SEC:Methods}.}
    \begin{NiceTabular}{clllllllll}
    \toprule
    &  Model    & Params & Epochs      & AP  & AP$_{50}$ & AP$_{75}$  & AP$_{\rm S}$ & AP$_{\rm M}$ & AP$_{\rm L}$ \\
    &           &        &             & (\%)  & (\%)  & (\%)  & (\%) & (\%) & (\%) \\
    \midrule
    \Block{3-1}{\rotate Bbox}
    & Gal-DETR            & 41M  & 500     & 22.6  & 38.1  & 26.2  & 16.3  &  24.8 & 19.8 \\
    & Gal-Deformable DETR & 40M  & 100     & 40.2  & 52.1  & 45.9  & 37.7  &  39.9 & 22.2 \\
    & Gal-DINO            & 47M  & 30      & 53.7  & 60.2  & 58.9  & 41.5  &  56.9 & 35.2 \\
    \midrule
    \Block{3-1}{\rotate Keys}
    & Gal-DETR            & 41M  & 500     & 35.4  & 37.5  & 35.3  & 9.1   &  60.0 & 49.6 \\
    & Gal-Deformable DETR & 40M  & 100     & 45.0  & 49.0  & 45.3  & 21.5  &  79.9 & 76.1 \\
    & Gal-DINO            & 47M  & 30      & 48.1  & 53.4  & 48.4  & 17.6  &  81.4 & 82.9 \\
    \bottomrule
    \end{NiceTabular}
    \label{TAB:AP1}
\end{table*}

\section{Training and Evaluation}
\label{SEC:Training}
In this section, we provide details of the network training process. 
Additionally, we outline the evaluation metrics employed for comparing different computer vision models in this section.

\subsection{Training Details}
The dataset division into training, validation, and test sets is detailed in Figure~\ref{FIG:Statistics} and Section~\ref{SEC:Datastats}.
The training data is used to train the networks, while the validation and test sets serve as inference datasets during and after training, respectively.
The networks described in Section~\ref{SEC:Methods} are trained for different numbers of epochs, which depend on their convergence speed and the stability of validation results. In this context, one epoch signifies a single pass through the entire training dataset during the model's training process.
During each epoch, the training dataset is subjected to various randomly applied augmentations.

These augmentations involve random flipping, random rotations, random resizing, and random cropping of a randomly selected set of 3-channel images from the training data.
Random flipping, performed horizontally (left to right), exposes the model to various object orientations and viewpoints. Random rotations, ranging from -180 to 180 degrees, promote rotational invariance in the network -- a critical aspect for handling radio galaxies with random orientations in the sky.
Random resizing operations entail scaling images up or down by selecting a target size randomly within predefined bounds. We have set the lower and upper bounds to $400\times400$ and $1300\times1300$ pixels, respectively. This strategy enables the model to learn from objects at different scales while preventing overfitting to specific sizes or aspect ratios.
Additionally, random cropping involves the random selection of a portion (or crop) from an input image, discarding the rest. This cropping operation, always applied after image resizing, introduces spatial variability in object locations within an image.
The application of these augmentations enhances the model's capacity to generalize effectively and perform well on unseen data.
These augmentations are applied during the training of all the networks described in this study.

All networks are trained and evaluated on an Nvidia Tesla P100.
We retained the original hyperparameters for each algorithm during training, except for specific modifications.
These hyperparameters are configuration settings that are not learned from the training data but are set prior to the training process.
The most important ones include the learning rate, which dictates the magnitude of weight adjustments during training; the batch size, indicating the number of data samples processed at a time; architecture parameters, specifying neural network architecture details like layer counts and filter sizes;  dropout rate for preventing overfitting by deactivating units during training; activation functions applied to the output of neural network layers to introduce non-linearity into the model; and optimizer for weight updates.
While we do not present all the hyperparameters for various networks here to keep the discussion concise, we refer the reader to the network architectures available in the provided repositories for these details.  
Here, we focus on the critical training aspects.

As detailed in Section~\ref{SEC:GalDETR}, Gal-DETR introduces multimodal modelling for radio and infrared detections.
Specifically, we implemented keypoint detection to the model architecture and Hungarian loss function. 
We reduced the learning rate to $5\times 10^{-5}$ and the number of queries to 10 from 100. 
The number of queries is decreased due to the fact that there are no more than five extended galaxies per image in our dataset, as depicted in the left panel of Figure~\ref{FIG:Statistics}.
As indicated in Table~\ref{TAB:AP1}, Gal-DETR has 41 million parameters and undergoes training over the course of 500 epochs for $\sim$10 hours on a single P100 GPU.

Similar changes were made for Gal-Deformable DETR model (Section~\ref{SEC:GaldDETR}), where keypoint detection was also implemented in the deformable attention mechanism. 
Training Gal-Deformable DETR, which comprises 40 million parameters, spanned 100 epochs and consumed approximately 8 hours on a single GPU.
For Gal-DINO model (Section~\ref{SEC:GalDINO}), we made the same changes as for Gal-DETR and additionally implemented keypoint detection in the de-noising anchor box mechanism.
Training Gal-DINO, featuring 47 million parameters, involved a 30-epoch training process that took approximately 6 hours on a single GPU.

As Gal-SIOD (Section~\ref{SEC:GalSIOD}) necessitates only one instance annotation per category within an image, we performed a selection process where we retained unique categories of radio galaxies in each image. 
For example, we ensured that each image contained at most one random annotation for FR-I sources, FR-II sources, FR-x sources, and R radio sources. 
This selection reduced the overall number of training annotations in all training images from 2,910 to 2,534, resulting in an annotation keeping ratio of 0.87.
No additional changes were made for Gal-SIOD-DMiner (Section~\ref{SEC:GalSIODD}) except for the reduction in annotations with the same keeping ratio.
Both the Gal-SIOD and Gal-SIOD-DMiner networks, each comprising 14.4 million parameters, underwent separate 200-epoch training sessions, with each session taking approximately 8 hours on a single GPU.
The Gal-Faster RCNN (Section~\ref{SEC:GalFRCNN}) model underwent training for 20,000 epochs, which took approximately 5 hours on a single GPU.
To train Gal-YOLOv8 (Section~\ref{SEC:GalYOLO}), we utilized the pre-trained YOLOv8 model from Ultralytics\footnote{https://ultralytics.com} for object detection. The training process took approximately 12 hours for 30 epochs and was conducted on a single GPU.

\begin{table*}
    \centering
    \caption{Bounding box detection results using Gal-SIOD and Gal-SIOD-DMiner networkd. The AP$_{50}$, AP$_{\rm S}$, AP$_{\rm M}$, and AP$_{\rm L}$ reported here correspond to those in Table~\ref{TAB:AP1}. The average precision values in this table are provided for various confidence thresholds ($S$), ranging from no limit to 0, 0.3, and 0.5 confidence scores. Comprehensive information regarding the models and their training (or evaluation) can be found in Sections~\ref{SEC:Methods} and \ref{SEC:Training}, respectively.}
    \begin{NiceTabular}{cccccccc}
    \toprule
Model & Confidence  & Params  & Epochs     & AP$_{50}$  & AP$_{\rm S}$ & AP$_{\rm M}$  & AP$_{\rm L}$  \\
      & Constraint  &         &            & (\%)    & (\%) & (\%) & (\%)  \\
    \midrule
    \Block{4-1}{Gal-SIOD}
    & AP$@S$        & 14.4M   & 200        & 24.0  & 14.7  & 17.8  & 14.8  \\
    & AP$@S_0$      & 14.4M   & 200        & 45.9  & 28.5  & 34.5  & 30.9  \\
    & AP$@S_3$      & 14.4M   & 200        & 41.6  & 26.6  & 31.4  & 22.4  \\
    & AP$@S_5$      & 14.4M   & 200        & 18.3  & 10.3  & 12.7  & 10.5  \\
    \midrule
    \Block{4-1}{Gal-SIOD-DMiner}
    & AP$@S$       & 14.4M   & 200         & 26.3  & 15.1  & 19.6  & 16.2  \\
    & AP$@S_0$     & 14.4M   & 200         & 46.8  & 27.3  & 35.4  & 30.3  \\
    & AP$@S_3$     & 14.4M   & 200         & 43.1  & 26.0  & 32.5  & 24.5  \\
    & AP$@S_5$     & 14.4M   & 200         & 24.3  & 15.6  & 16.3  & 12.1  \\
    \bottomrule
    \end{NiceTabular}
    \label{TAB:AP2}
\end{table*}

\begin{table}
    \centering
    \caption{Bounding box detection results for Gal-Faster RCNN and Gal-YOLOv8 models. The columns here align with those presented in Table~\ref{TAB:AP1}. Additional information regarding the networks can be found in Sections~\ref{SEC:Methods} and \ref{SEC:Training}.}
    \begin{tabular}{lcccccc}
    \toprule
Model           & AP    & AP$_{50}$ &AP$_{75}$ & AP$_{\rm S}$ & AP$_{\rm M}$    & AP$_{\rm L}$ \\
                & (\%)  & (\%)      &  (\%)    & (\%)         & (\%) & (\%) \\
\midrule
Gal-Faster RCNN  & 31.9  & 48.4      & 37.9     & 26.6         & 33.3     & 22.9\\
Gal-YOLOv8      & 54.5  & 58.5      & 57.2     & 45.5         & 55.4     & 34.0\\
    \bottomrule
    \end{tabular}
    \label{TAB:AP3}
\end{table}

\subsection{Evaluation Metric}
We employ the standard evaluation metrics outlined by \cite{coco14} for computer vision to evaluate and compare the performance of the seven algorithms on the test dataset. 
Specifically, we utilize the Intersection over Union (IoU), a metric commonly used for assessing object detection algorithm performance. 
The IoU is calculated as the ratio of the intersection between the predicted bounding boxes ($P_{\rm B}$) and the ground truth bounding boxes ($T_{\rm B}$) to the union of the two boxes. The metric can be expressed as
\begin{equation}
    \text{IoU}(P_{\rm B}, T_{\rm B}) = \frac{{\text{Area of Overlap between $P_{\rm B}$ and $T_{\rm B}$}}}{{\text{Area of Union between $P_{\rm B}$ and $T_{\rm B}$}}},
\end{equation}
In the domain of bounding box prediction, the area of each predicted box is evaluated to ascertain its classification as a true positive (TP), false positive (FP), or false negative (FN) concerning the ground truth box's area for each radio galaxy.
To be more specific, a bounding box is classified as TP if it accurately identifies an object and exhibits a sufficiently high IoU overlap with a corresponding ground truth bounding box. 
In essence, this means that the algorithm's detected bounding box correctly recognizes an object present in the ground truth data. 
An FP bounding box, on the other hand, is one generated by the algorithm but fails to accurately correspond to any ground truth object. 
Finally, a false negative (FN) bounding box occurs when an object is present in the ground truth data, but the algorithm fails to detect it, indicating a missed opportunity to identify a genuine object.

The metric employed to assess the keypoint detection of infrared hosts is called Object Keypoint Detection (OKS). It is determined by the Euclidean distance between the predicted and true positions of the infrared hosts and is expressed as:
\begin{equation}
\mathrm{OKS} = \exp \left(-\frac{d^2}{2r^2c^2}\right).
\end{equation}
In this equation, $d$ represents the Euclidean distance, $r$ represents the ratio of the bounding box to the image area, and $c=0.107$ instead of 10 values in \cite{coco14}, as there is one infrared host per bounding box in our case. 
The OKS score ranges from 0 to 1, with 1 signifying optimal detection.

We assess the performance of all seven networks using the average precision metric, a widely adopted standard for evaluating object detection models \citep[][]{coco14}. 
Precision is a metric that measures the accuracy of positive predictions made by a model. It is calculated as the ratio of TPs (correctly identified instances of the positive class) to the sum of TPs and FPs (incorrectly identified instances of the positive class). 
Recall, also known as sensitivity or true positive rate, is a metric that measures the ability of a model to correctly identify all relevant instances of the positive class. It is calculated as the ratio of TPs to the sum of TPs and FPs (instances of the positive class that were not identified).  
The Precision-Recall Curve depicts how the balance between precision and recall changes at various thresholds.
Average Precision (AP) is a metric calculated from the precision-recall curve, representing the area under that curve. 
An AP of 1 indicates a perfect model, achieving both maximum precision and recall. 
On the other hand, an AP of 0 signifies that the model fails to correctly identify any positive instances, resulting in minimal precision and recall.

\begin{figure*}
\centering
\includegraphics[width=8.5cm, scale=0.5]{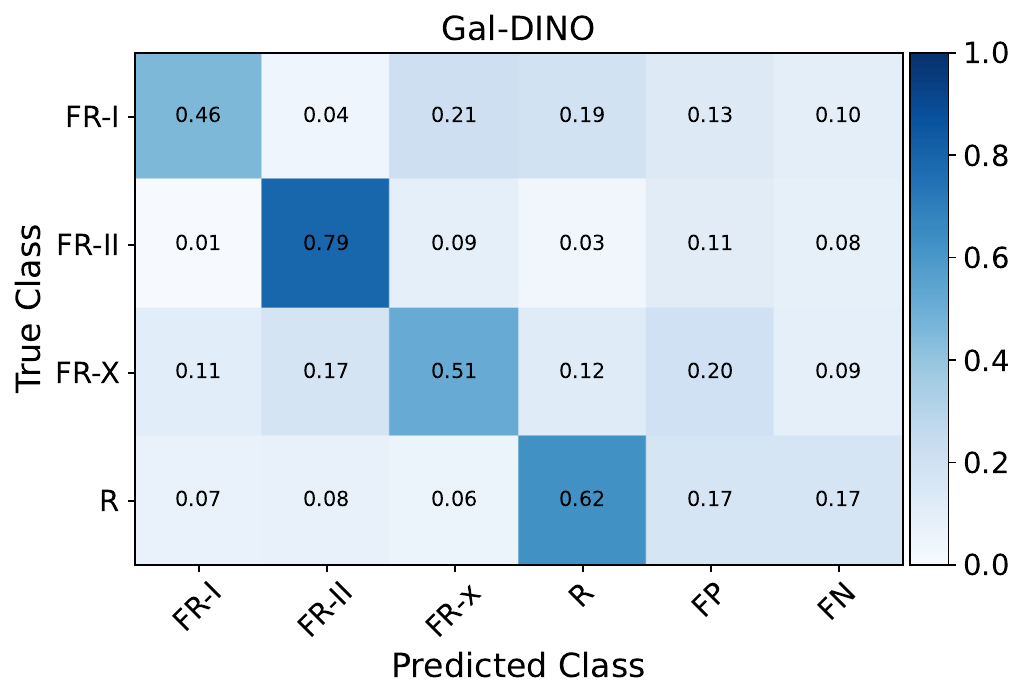}
\includegraphics[width=8.5cm, scale=0.5]{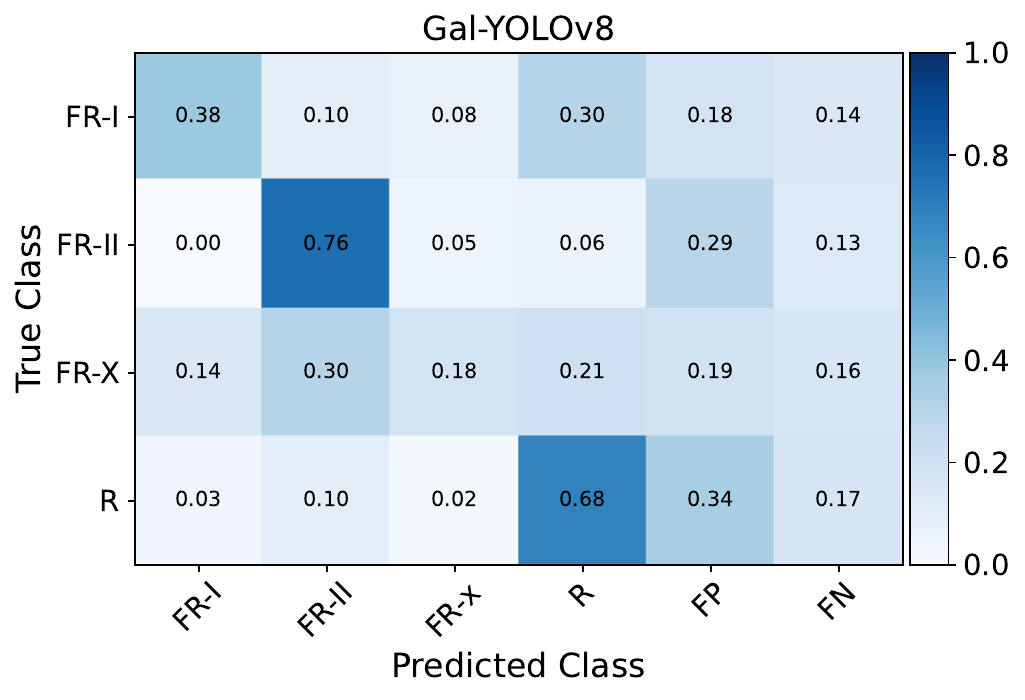}
\caption{Confusion Matrices: Shown are the normalized matrices for the Gal-DINO model and Gal-YOLOv8 detection model in the left and right panels, respectively. 
Here the diagonal values corresponding to various galaxy classes represent the true positive (TP) instances at IoU and confidence thresholds of 0.5 and 0.25, respectively. Beyond these thresholds, the false positive (FP) values indicate detections without corresponding ground truth instances, while the false negative (FN) values signify instances where the model failed to detect the galaxies.
} 
\label{FIG:Confusion}
\end{figure*}

We compute average precision values by employing standard IoU and OKS thresholds for bounding boxes and keypoints, respectively. 
The IoU threshold determines the correctness of a prediction based on the extent of overlap it shares with the ground truth; predictions with an IoU exceeding this threshold are considered correct. 
Similarly, the OKS threshold assesses the correctness of predicted keypoints based on their similarity scores, with scores above this threshold indicating correctness.
We calculate average precision at IoU (or OKS) thresholds ranging from 0.50 to 0.95, denoted as AP, as well as at specific IoU (or OKS) thresholds of 0.50 (AP$_{50}$) and 0.75 (AP$_{75}$) for radio galaxies of all sizes. 
Additionally, in order to assess performance across varying scales of structures within radio images, we compute the Average Precision for small (AP$_{\rm S}$), medium (AP$_{\rm M}$), and large (AP$_{\rm L}$) area ranges, defined as pixel areas less than $24^2$, between $24^2$ and $48^2$, and greater than $48^2$, respectively.

We evaluate the performance of bounding boxes for radio galaxies and the identification of keypoints indicating the positions of infrared hosts using the multimodal networks Gal-DINO, Gal-Deformable DETR, and Gal-DINO, all introduced in the present study.
For Gal-SIOD, Gal-SIOD-DMiner, Gal-Faster RCNN, and Gal-YOLOv8 networks, we exclusively provide average precision values for the bounding boxes. 
In the case of Gal-SIOD and Gal-SIOD-DMiner, a confidence constraint to the COCO-style evaluation metrics is used. 
This constraint requires a predicted box to meet specific IoU and confidence threshold ($S$) criteria in order to be considered a true match.

\section{Results}
\label{SEC:Results}
We evaluate the performance of state-of-the-art object detection algorithms on our dataset, which includes processed radio and infrared images and annotations that are tailored for the detection of extended radio galaxies. 
Experiments involving noisy radio are discussed in \ref{SEC:Noisy}. 
Our novel multimodal modelling approach is introduced to enable the simultaneous detection of radio galaxies and their corresponding infrared hosts. 
This approach incorporates keypoint detection into the Gal-DETR, Gal-Deformable DETR, and Gal-DINO algorithms. Additionally, we train Gal-SIOD, Gal-SIOD-DMiner, Gal-Faster RCNN, and Gal-YOLOv8 models to compare their performance in radio galaxy bounding box detection.

Table~\ref{TAB:AP1} provides the Gal-DETR, Gal-Deformable DETR, and Gal-DINO results for both bounding box and keypoint detection on the test set of the RadioGalaxyNET dataset. 
Notably, Gal-DINO emerged as the top-performing model, achieving the highest AP of 53.7\%. 
This outstanding performance underscores its excellence in detecting bounding boxes for radio galaxies. 
Gal-Deformable DETR also demonstrated competence in identifying radio galaxies, with an AP of 40.2\%. 
In contrast, Gal-DETR, while capable, achieved a comparatively lower AP of 22.6\%. Gal-DINO's superiority extended across all IoU thresholds (AP$_{50}$ and AP$_{75}$), securing an AP$_{50}$ of 60.2\% and an AP$_{75}$ of 58.9\%, signifying its robustness in radio galaxy detection across varying IoU thresholds. 
Moreover, Gal-DINO consistently outperformed other models for small, medium, and large-sized radio galaxies (AP$_{\rm S}$, AP$_{\rm M}$, and AP$_{\rm L}$). Additionally, Gal-DINO excelled in keypoint detection, achieving an AP of 48.1\% and an AP$_{50}$ of 53.4\%. 
In summary, these results affirm Gal-DINO as the most effective model for both bounding box and keypoint detection in the realm of radio galaxy and infrared host detection, displaying strong performance across diverse IoU thresholds and galaxy sizes.

Table~\ref{TAB:AP2} presents the outcomes for Gal-SIOD and Gal-SIOD-DMiner, incorporating a confidence constraint into the COCO-style evaluation metrics. A predicted box is considered a true match only when it meets particular IoU and confidence threshold ($S$) criteria. 
Here also we use the same area range as above for evaluation.
With no confidence constraint ($S=0$), Gal-SIOD achieves a relatively high AP$_{50}$ of 45.9\%. 
As the confidence threshold increases, the overall AP values decrease. For instance, at $S=0.3$ and $S=0.5$, the AP$_{50}$ drops to 41.6\% and 18.3\%, respectively.
The model exhibits similar trends across all confidence thresholds for small, medium, and large-sized radio galaxies, as seen in the values of AP$_{\rm S}$, AP$_{\rm M}$, and AP$_{\rm L}$.
Gal-SIOD-DMiner achieves slighly higher AP$_{50}$ of 46.8\% with no confidence constraint.
Similar to Gal-SIOD, as the confidence threshold increases, the overall AP values decrease. 
The model also exhibits consistent performance trends across all confidence thresholds for small, medium, and large-sized radio galaxies.

Figure~\ref{FIG:Results} displays RGB images and ground truth annotations (first column), ground truth and predicted keypoints as circles and triangles over infrared images (second column), Gal-DINO bounding box predictions over radio images (third column), and Gal-SIOD-DMiner predictions over radio images (fourth column). All predictions are above the confidence threshold of 0.25. 
Notably, these models detect more extended galaxies where the ground truth annotation is not available in the same frame.
However, it is worth noting that some of these detections do not correspond to FR radio galaxies; instead, they include peculiar radio sources like the Odd Radio Circle, as illustrated in the bottom right panel of the figure.
Table ~\ref{TAB:AP3} shows results for Gal-Faster RCNN and Gal-YOLOv8 models. 
Gal-YOLOv8 outperforms Gal-Faster RCNN by a significant margin, achieving an AP$_{50}$ of 58.5\% compared to 48.4\% for Gal-Faster RCNN. 
Gal-YOLOv8 and Gal-DINO are among the best performing models that exhibit similar AP of 54.5\% and 53.7\%, respectively. 
Both models achieve high AP scores, with Gal-YOLOv8 displaying a slight advantage in detecting smaller radio galaxies, while Gal-DINO excels in identifying medium and large-sized galaxies.

Figure~\ref{FIG:Confusion} shows confusion matrices for the best-performing Gal-DINO and Gal-YOLOv8 models. 
These matrices are calculated for IoU (and OKS for Gal-DINO) and confidence thresholds of 0.5 and 0.25, respectively.
In object detection, the confusion matrix, as explained in \citet{gupta2023a}, differs from the one used in classification. 
Unlike classification, where each image has only one label, object detection allows for multiple instances of the same or different classes within an image. 
This leads to multiple TP, FP, and FN values for each class in the confusion matrix. 
When constructing the confusion matrix, attention is paid to the intersection between predicted and true bounding boxes (and keypoints for Gal-DINO), assessed using an IoU (and OKS for Gal-DINO) threshold of 0.5.
The left panel of Figure~\ref{FIG:Confusion} displays Gal-DINO model performance. In the FR-II class, the TP value indicates 79\% accurate detection in the test set. However, there are moderate FPs, signifying 11\% mispredictions, and FNs, suggesting an 8\% miss rate. Similar patterns are observed for the FR-I, FR-x, and R classes.
The right panel shows that Gal-YOLOv8 displays a behaviour similar to that of Gal-DINO. However, the TP value for FR-x is considerably smaller, indicating that the model tends to confuse FR-II or FR-I sources with FR-x to a greater extent.

\section{Conclusions}
\label{SEC:Conclusions}
The next generation of radio telescopes has the capability to rapidly survey extensive regions of the sky, culminating in the generation of extensive catalogues comprising millions of radio galaxies.
Nevertheless, these highly sensitive surveys are also detecting a very large number of complex radio sources, rendering conventional source extraction approaches less useful.
We introduce RadioGalaxyNET, comprising a labelled dataset and a suite of computer vision models designed for detecting extended radio sources and their corresponding infrared host galaxies. 
The dataset encompasses 2,800 images, comprising both radio and infrared sky channels, and includes 4,155 instances of labelled sources. 
Radio images are sourced from the Evolutionary Map of the Universe pilot survey (EMU-PS), which was carried out using the Australian Square Kilometre Array Pathfinder (ASKAP) telescope. 
The identification of radio galaxies and the curation of annotations, including class labels, bounding boxes, and segmentation masks, are performed through visual inspections. 
Additionally, infrared images from the Wide-field Infrared Survey Explorer (WISE) are aligned with the sky positions of the radio images, and the infrared host galaxies are visually identified in the infrared images.

Our computer vision methods include multimodal models, including Gal-DETR, Gal-Deformable DETR, and Gal-DINO, where we introduce a mechanism for simultaneous detection of radio galaxies and infrared hosts through bounding box and keypoint detections, respectively.
We employ the Average Precision (AP) metric to assess all the models presented in this study for both the prediction of bounding boxes for radio sources and the prediction of keypoints for infrared hosts.
Our findings reveal that the Gal-DINO model exhibits superior performance in detecting both radio galaxies and infrared hosts, achieving AP$_{50}$ scores of 60.2\% and 53.4\%, respectively. 
In addition to these models, we conduct a comparative analysis of our radio galaxy detection results with those produced by Gal-SIOD, Gal-SIOD-DMiner, Gal-Faster RCNN, and Gal-YOLOv8 models. Our results indicate that Gal-DINO and Gal-YOLOv8 yield comparable outcomes in radio galaxy detection.  Future work should consider our dataset and the computer vision methodologies for cataloguing radio galaxies and infrared hosts in ongoing and upcoming radio surveys.
The availability of our dataset will facilitate the development of more advanced machine learning techniques for detecting radio galaxies and infrared hosts in the next generation of radio surveys, such as the EMU main survey.
In order to further enrich the diversity of training data across the entire southern sky, future research should also explore the possibilities of incorporating online learning and active learning with a human-in-the-loop approach for the EMU main survey.

\section*{Data Availability}

RadioGalaxyNET dataset can be downloaded from \url{https://doi.org/10.25919/btk3-vx79}.
Network architectures for Gal-DETR, Gal-Deformable DETR and Gal-DINO can be cloned from \url{http://hdl.handle.net/102.100.100/602494?index=1}.
GitHub to all networks:
\begin{itemize}
    \item Gal-DETR: \url{https://github.com/Nikhel1/Gal-DETR}
    \item Gal-Deformable DETR: \url{http://github.com/Nikhel1/Gal-Deformable-DETR}
    \item Gal-DINO: \url{https://github.com/Nikhel1/Gal-DINO}
    \item Gal-SIOD and Gal-SIOD-DMiner: \url{https://github.com/Nikhel1/Gal-SIOD}
    \item Gal-Faster~RCNN:~\url{https://github.com/Nikhel1/Gal-Faster-RCNN}
    \item Gal-YOLOv8: \url{https://github.com/Nikhel1/Gal-YOLOv8}
\end{itemize}

\section*{Acknowledgements}
The Australian SKA Pathfinder is part of the Australia Telescope National Facility, which is managed by CSIRO. The operation of ASKAP is funded by the Australian Government with support from the National Collaborative Research Infrastructure Strategy. ASKAP uses the resources of the Pawsey Supercomputing Centre. The establishment of ASKAP, the Murchison Radio-astronomy Observatory and the Pawsey Supercomputing Centre are initiatives of the Australian Government, with support from the Government of Western Australia and the Science and Industry Endowment Fund. We acknowledge the Wajarri Yamatji people as the traditional owners of the Observatory site.
The photometric redshifts for the Legacy Surveys (PRLS) catalogue used in this paper were produced thanks to funding from the U.S. Department of Energy Office of Science and Office of High Energy Physics via grant DE-SC0007914.
This research has made use of the NASA/IPAC Extragalactic Database (NED), which is operated by the Jet Propulsion Laboratory, California Institute of Technology, under contract with the National Aeronautics and Space Administration.
NG acknowledges support from CSIRO’s Machine Learning and Artificial Intelligence Future Science (MLAI FSP) Platform.

\bibliography{ASKAP_PASA}

\appendix

\section{Model Analysis with Noisy/Raw Radio Images}
\label{SEC:Noisy}

\begin{table}
    \centering
    \caption{Bounding box and keypoint detection results on the test set of RadioGalaxyNET. Instead of using processed images, the 3-channel RGB images used for training and testing the networks include two channels that contain noisy raw radio information, and one channel has processed infrared images.}
    \begin{NiceTabular}{lllllll}
    \toprule
    &  Model        & AP  & AP$_{50}$ & AP$_{75}$  & AP$_{\rm S}$ & AP$_{\rm M}$  \\
    &                       & (\%)  & (\%)  & (\%)  & (\%) & (\%)  \\
    \midrule
    \Block{3-1}{\rotate Bbox}
    & Gal-DETR               & 12.5  & 32.0  & 5.2   &  7.2   &  14.2  \\
    & Gal-Deform. DETR     & 15.7  & 33.1  & 11.7  & 11.3  &  16.9  \\
    & Gal-DINO-4scale           & 23.0  & 47.3  & 16.2  & 23.3  &  21.5 \\
    \midrule
    \Block{3-1}{\rotate Keys}
    & Gal-DETR                 & 32.5  & 33.6  & 32.3  &  6.3   &  55.0 \\
    & Gal-Deform. DETR     & 33.8  & 34.5  & 33.8  & 14.8  &  72.1  \\
    & Gal-DINO-4scale         & 43.1  & 47.1  & 44.5  & 22.1  &  76.5  \\
    \bottomrule
    \end{NiceTabular}
    \label{TAB:AP1_sup}
\end{table}

\begin{figure*}
\centering
\includegraphics[width=13cm, scale=0.5]{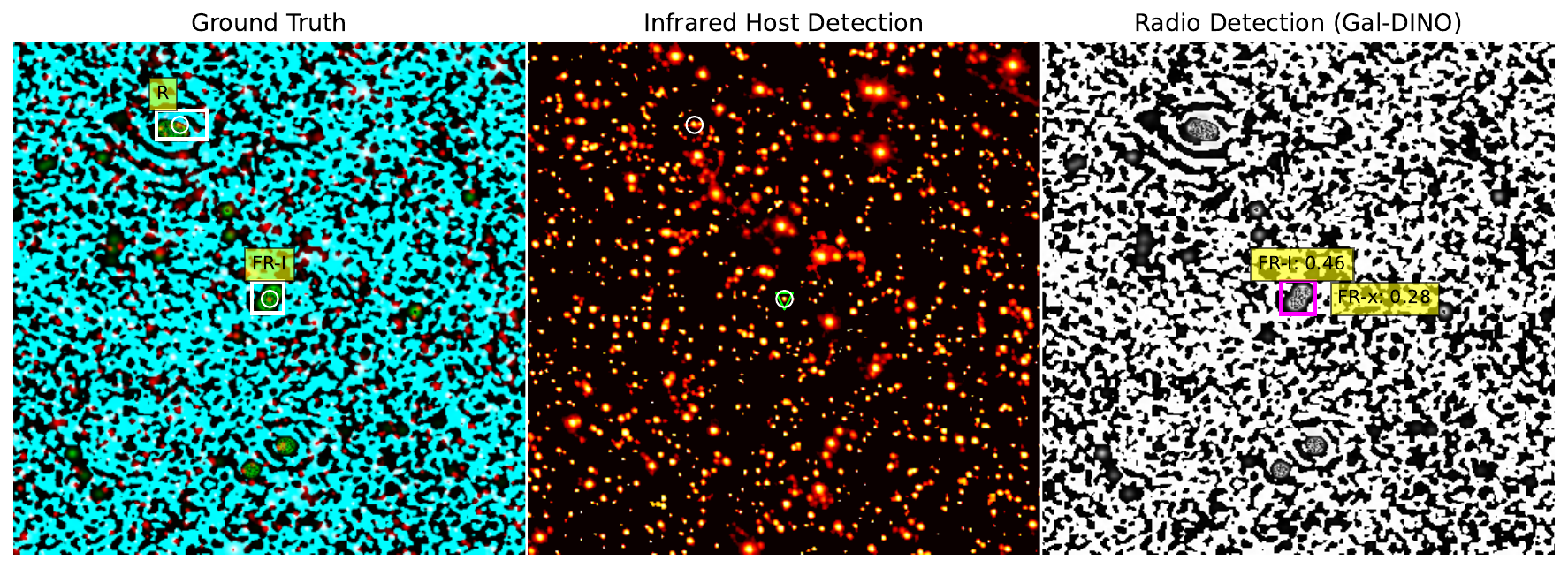}
\includegraphics[trim={0cm 0cm 0cm 1.25cm}, clip, width=13cm, scale=0.5]{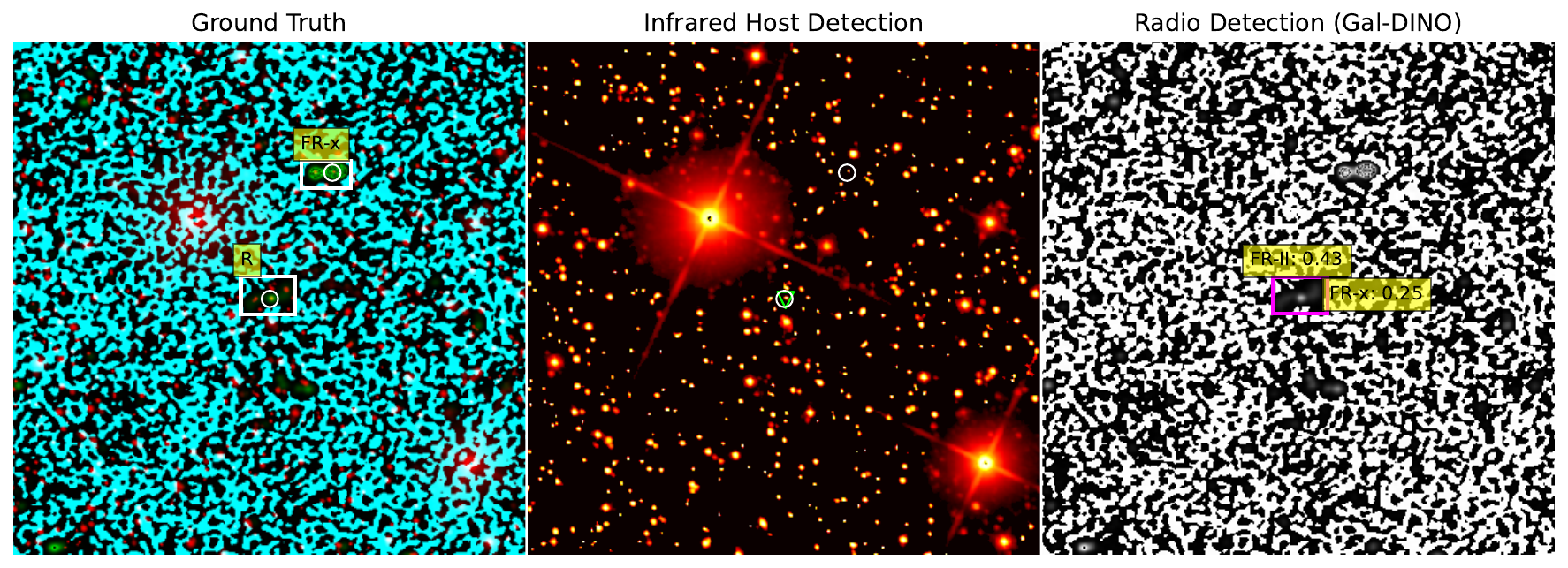}
\includegraphics[trim={0cm 0cm 0cm 1.25cm}, clip, width=13cm, scale=0.5]{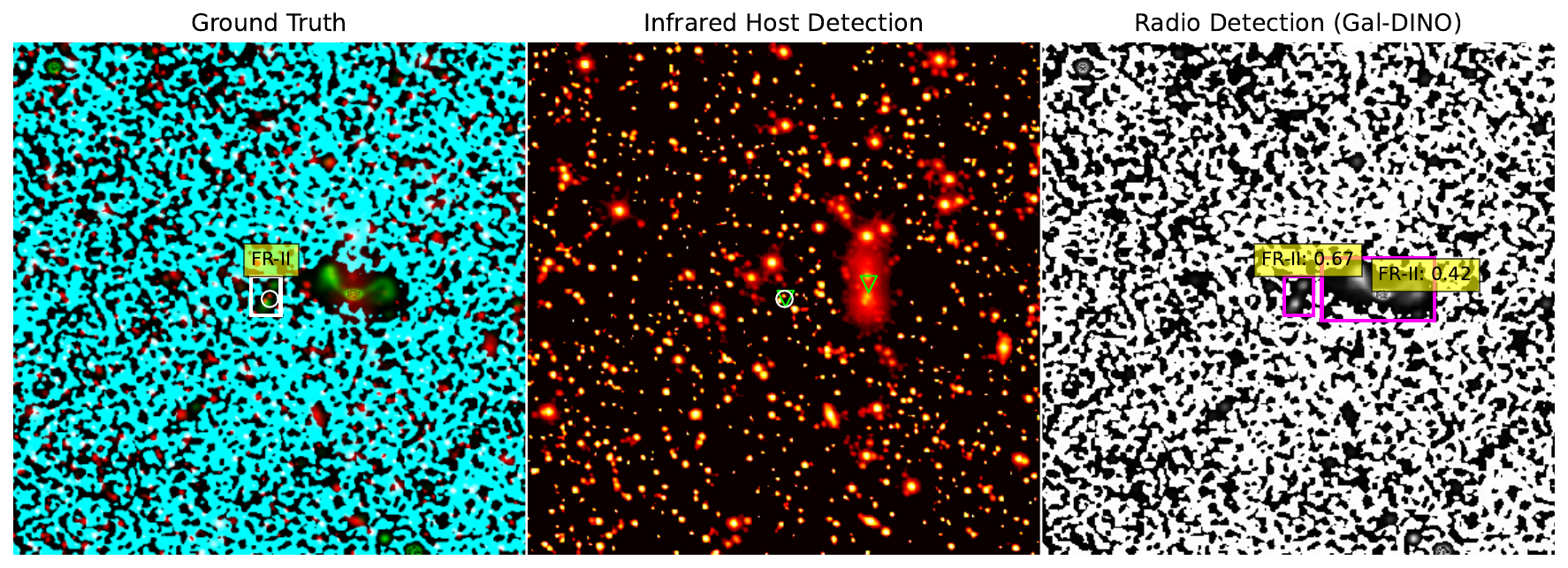}
\includegraphics[trim={0cm 0cm 0cm 1.25cm}, clip, width=13cm, scale=0.5]{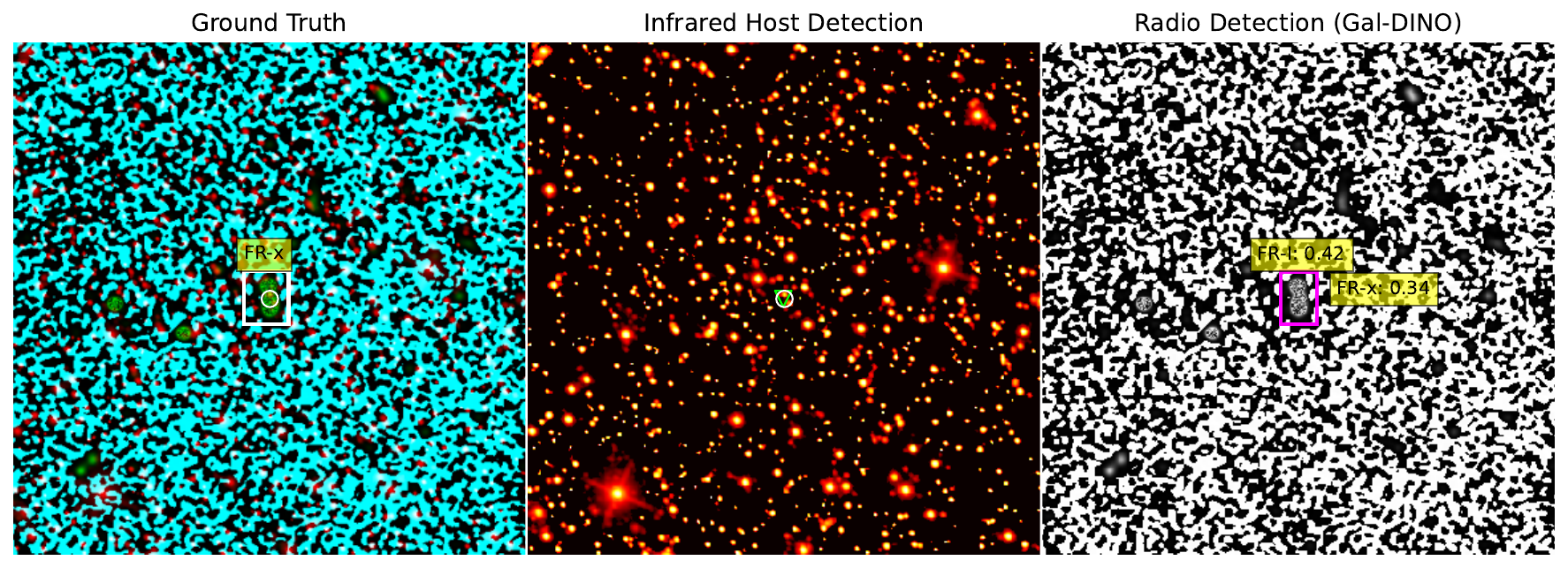}
\includegraphics[trim={0cm 0cm 0cm 1.25cm}, clip, width=13cm, scale=0.5]{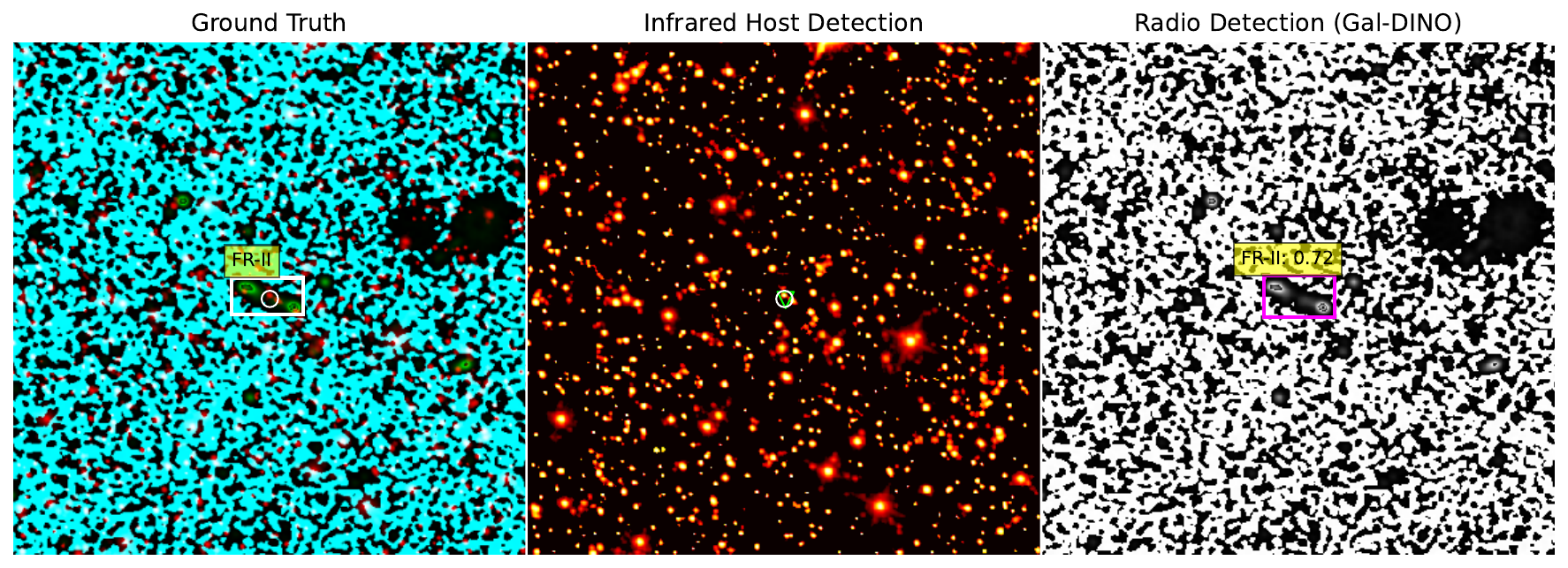}
\caption{Object detection results: Shown are the raw radio and processed infrared 3-channel RGB images and ground truth annotations (first column), ground truth and Gal-DINO keypoint detections as circles and triangles over infrared images (second column), Gal-DINO class and bounding box predictions over raw radio images (third column). Better viewed in colour.} 
\label{FIG:Results_sup}
\end{figure*}

We provide object detection results using raw radio images in this additional study.
The 3-channel RGB images utilized for training and testing the networks consist of two channels containing noisy raw radio images instead of processed ones. 
Specifically, the B channel is filled with 8-16 bit raw radio information, while the G channel contains 0-8 bit raw radio information.
For the R channel, we keep 8-16 bit infrared information as before.
We train Gal-DETR, Gal-Deformable DETR and Gal-DINO-4scale multimodel models using the same set of hyperparameters and annotations as before.
In Figure~\ref{FIG:Results_sup}, the first column showcases RGB images with ground truth annotations. The second column displays the ground truth and predicted keypoints represented by circles and triangles, respectively, overlaid on infrared images. The third column shows the bounding box predictions made by the Gal-DINO model over radio images. All predictions shown in the figure are above the confidence threshold of 0.25.

Table~\ref{TAB:AP1_sup} showcases the evaluation outcomes of Gal-DETR, Gal-Deformable DETR, and Gal-DINO for the detection of bounding boxes around extended radio galaxies and keypoints representing the positions of infrared host galaxies. 
It is worth noting that the Average Precision values are lower compared to the results achieved with processed radio images, as presented in the main paper. 
The presence of noise in radio images poses additional challenges to object detection tasks. 
Nevertheless, with further enhancements, these models have the potential to be directly applied to raw radio data, expanding their applicability.

\label{lastpage}
\end{document}